%% file: main.tex
\documentclass[conference,compsoc]{IEEEtran}

\input{header}

\ifCLASSOPTIONcompsoc
  \usepackage[nocompress]{cite}
\else
  \usepackage{cite}
\fi

\ifCLASSINFOpdf
\else
\fi

\begin{document}

\title{DeepInvert: Semi-Supervised Embedding Inversion \\ Against Obfuscated Language Models}


\author{\IEEEauthorblockN{Zhicong Huang}
\IEEEauthorblockA{Ant Group
}
\and
\IEEEauthorblockN{Cheng Hong}
\IEEEauthorblockA{Ant Group
}
\and
\IEEEauthorblockN{Tao Wei}
\IEEEauthorblockA{Ant Group
}}

\newcommand{\cheng}[1]{\textcolor{red}{Cheng: {#1}}}

\maketitle

\begin{abstract}
  Cloud-based language model services routinely process prompts containing sensitive personal or business information. Obfuscation-based defenses---including ObfusLM (ACL 2025), SentinelLMs (AAAI 2024), TextObfuscator (ACL 2023), and DPNR (EMNLP 2020)---mitigate this risk by transforming prompt representations before transmission, offering a lightweight alternative to heavyweight cryptographic solutions. We show that these defenses provide far less protection than previously believed.

  We present \NAME, a semi-supervised embedding inversion attack that recovers original tokens from obfuscated representations with substantially higher accuracy than prior methods. The key insight is that unlabeled obfuscated embeddings, despite perturbation, retain exploitable semantic structure. \NAME~combines supervised training on labeled shadow data with a novel unsupervised consistency objective over unlabeled target embeddings, and alternates between the two via a mixed training pipeline that enables mutual reinforcement. Defense-aware adaptations further extend the attack to diverse obfuscation mechanisms across both encoder-based and autoregressive architectures.

  Experiments on nine defenses, five tasks, and four model architectures show that \NAME~outperforms prior attacks on most defenses and tasks. Against the state-of-the-art ObfusLM, \NAME~achieves 73.5\% top-1 token recovery versus 26.2\% for the previous best attack. On token-level and generation tasks, our results reveal a \revised{task-dependent} tension: obfuscation schemes that preserve enough signal for downstream utility also retain sufficient structure for accurate inversion, while schemes that resist inversion collapse task utility. \revised{On simpler sentence-level classification, some DP-based defenses can simultaneously maintain utility and limit recovery, suggesting the tension is task-dependent.} We call for a re-evaluation of this class of defenses.
\end{abstract}

\IEEEpeerreviewmaketitle

\input{sections/intro}
\input{sections/background}
\input{sections/design}

\input{sections/attack}
\input{sections/evaluation}
\input{sections/related}
\input{sections/conclusion}




\bibliographystyle{ieeetr}
\bibliography{ref}


\input{sections/appendix}

\end{document}

%% file: header.tex
\PassOptionsToPackage{table}{xcolor}
\usepackage[normalem]{ulem}
\usepackage{amsmath}
\usepackage{amsthm}
\usepackage{fvextra}
\usepackage{framed}
\usepackage{stmaryrd}
\usepackage{tikz}
\usetikzlibrary{fit}

\usepackage[linesnumbered,ruled]{algorithm2e}
\usepackage{graphicx}
\usepackage[font=footnotesize]{subfig}
\usepackage{tikz}
\usepackage{array}
\usepackage{eqparbox}
\usepackage{epstopdf}
\usepackage{pgfplots}
\usepackage{textcomp}
\usepackage{multirow}
\usepackage{xcolor}
\usepackage{enumitem}
\usepackage{balance}
\usepackage{booktabs}
\usepackage{pifont}
\usepackage{url}
\usepackage{siunitx}
\usepackage{threeparttable}
\usepackage{makecell}
\usepackage{comment}
\usepackage{verbatim}

\usepackage{amssymb,amsfonts}

\usepackage{tikz}
\usetikzlibrary{calc}

\newlength{\commentindent}

\newcommand{\tNAME}{\mbox{DeepInvert}}
\newcommand{\NAME}{\textit{\tNAME}}

\newcommand{\RR}{\mathbb{R}}
\newcommand{\EE}{\mathbb{E}}

\newcommand{\red}[1]{{\color{red} #1}}
\newcommand{\orange}[1]{{\color{orange} #1}}
\newcommand{\cyan}[1]{{\color{cyan} #1}}

\newcommand{\highlight}[1]{{\color{blue} #1}}
\newcommand{\sensitive}[1]{{\color{red} #1}}

\newcommand{\revised}[1]{#1}

\newtheorem{lemma}{Lemma}

\newcommand{\best}[1]{\cellcolor{gray!20}\textbf{#1}}

\newcommand{\scoreColorCell}[2]{%
  \begingroup
  \ifdim #1pt < 0.10pt \cellcolor[HTML]{FFFDFB}\else
  \ifdim #1pt < 0.30pt \cellcolor[HTML]{FFF7F0}\else
  \ifdim #1pt < 0.50pt \cellcolor[HTML]{FEEEDD}\else
  \ifdim #1pt < 0.70pt \cellcolor[HTML]{FDE2C8}\else
  \ifdim #1pt < 0.85pt \cellcolor[HTML]{FBD4AE}\else
  \cellcolor[HTML]{F7C08A}\fi\fi\fi\fi\fi
  #2%
  \endgroup
}
\newcommand{\score}[1]{\scoreColorCell{#1}{#1}}
\renewcommand{\best}[1]{\scoreColorCell{#1}{\textbf{#1}}}

\newcommand{\fin}{\mathbb{R}^{L\times H}}
\newcommand{\fout}{\mathbb{R}^{L\times V}}

\newcommand{\kl}{\text{KL}}
\newcommand{\mask}{[\mathsf{MASK]}}
\newcommand{\loss}{\mathcal{L}}

\SetCommentSty{normalfont}
\newcommand{\mytcp}[1]{\tcp{\textbf{#1}}}

%% file: sections/intro.tex
\section{Introduction}
Recent advances in language models (LMs)—from encoder-based models such as BERT~\cite{DBLP:conf/naacl/DevlinCLT19} to large decoder-only models such as OpenAI's GPT family~\cite{achiam2023gpt}, Meta's Llama series~\cite{touvron2023llama, grattafiori2024llama}, and Alibaba's Qwen series~\cite{DBLP:journals/corr/abs-2412-15115}—have delivered state-of-the-art performance across a wide range of NLP tasks. However, these models are often extremely large, computationally demanding, or partially closed-source, which makes self-hosted deployment impractical for many users. As a result, LM functionality is frequently accessed via cloud services, where clients either upload datasets to adapt a model to a downstream task (\emph{fine-tuning}) or send prompts to query an existing model (\emph{inference}). This cloud-based paradigm raises significant privacy and security concerns~\cite{DBLP:conf/ccs/SongR20, DBLP:conf/sp/PanZJY20}. Client inputs, either fine-tuning data or inference-time prompts, may contain sensitive personal information, proprietary business data, or other confidential content that should not be exposed to the service provider or potential attackers with access to the provider's infrastructure.

A series of works propose mechanisms to mitigate these risks. Obfuscation-based defenses are particularly attractive because they can be integrated into existing pipelines with minimal changes and relatively low computational overhead. Some methods rely on heuristic techniques (e.g., vocabulary permutation~\cite{DBLP:conf/aaai/MishraLD24, DBLP:conf/acl/LinYMZH0WLCD025}, data mixing~\cite{DBLP:conf/eccv/LiuWGZH20}), while others aim for more principled designs with formal privacy guarantees, most commonly via differential privacy (DP)~\cite{DBLP:conf/acl/YueDWLSC21, DBLP:conf/emnlp/LyuHL20, DBLP:conf/ccs/DuYC0H023}. The core idea is to transform or perturb prompt representations on the client side before transmitting them to the server. A well-known limitation of these approaches is the inevitable degradation in utility caused by perturbation. However, as we show in this work, the more critical issue is that \textit{many proposed defenses can be effectively broken under standard, recommended privacy parameters} (we revisit this point in Section~\ref{sec:discussion}). An alternative research line seeks stronger protection from heavyweight cryptographic primitives that support computation over encrypted data—such as homomorphic encryption and secure multiparty computation~\cite{DBLP:conf/ndss/ZhangYH0LWH00025, DBLP:conf/ccs/MoonYJK25, DBLP:conf/ndss/LuHGL000WC25, DBLP:conf/sp/PangZMZS24}—but these often incur substantial computational and communication costs.

\input{tables/atk_examples_llama_obfuslm.tex}

Prior work has shown that information can be recovered from obfuscation-based defenses, but existing attacks remain relatively shallow and not consistently effective. For instance, the KNN attack~\cite{DBLP:conf/cikm/QuKY0BN21} demonstrates that directly computing similarities between obfuscated embeddings and public word embeddings can re-identify a fraction of tokens, yet some defenses~\cite{DBLP:conf/acl/ZhouLMGWDZZH23, DBLP:conf/acl/LinYMZH0WLCD025} can largely thwart this approach. The InvBert attack~\cite{DBLP:journals/corr/abs-2109-10104} trains a model to predict token IDs from obfuscated embeddings, but its fully supervised training paradigm cannot exploit the rich information contained in \emph{unlabeled} obfuscated embeddings. We close this gap with a semi-supervised attack that allows the adversary to leverage both labeled and unlabeled obfuscated embeddings, recover finer-grained semantic structure, and thereby expose a markedly larger fraction of the original tokens in obfuscation-based defenses.

 \noindent \textbf{\textit{Contributions.}}
 \revised{We propose \NAME, a semi-supervised embedding inversion attack that, to our knowledge, is the first to exploit unlabeled target embeddings for prompt recovery—a capability that prior supervised or optimization-based attacks lack.} The core innovation is a consistency-based unsupervised objective that extracts invariant semantic structure from noisy obfuscated embeddings, combined with a mixed training pipeline that alternates supervised and unsupervised optimization for mutual reinforcement. While individual building blocks (pseudo-labeling, EMA teachers) originate from semi-supervised learning, applying them to embedding inversion poses non-trivial challenges—large-vocabulary prediction, defense-induced distribution shift, and defense-specific noise characteristics—that require tailored solutions. Our main contributions are:

\begin{itemize}
\item \NAME\ improves token recovery by extracting deeper semantic structure from both labeled and unlabeled obfuscated embeddings, addressing the overfitting to the shadow distribution that limits prior supervised attacks.
\item We identify and validate specific adaptations critical for inversion: soft top-$k$ labels, masked augmented views at the embedding level, and an EMA teacher for stable self-training.
\item We tailor \NAME\ to different defenses via multiple shadow models (to mitigate distribution shift) and renormalization/denoising (for DP-based defenses).
\item We systematically evaluate \NAME\ on nine defenses, five tasks, and four model architectures. \NAME\ breaks ObfusLM with 73.5\% top-1 recovery (vs.\ 26.2\% for the previous best). \revised{Our results reveal a task-dependent tension: for token-level and generation tasks, current schemes fail to achieve meaningful privacy-utility trade-off, while some DP-based defenses can maintain both on simpler sentence-level tasks.}
\end{itemize}

%% file: tables/atk_examples_llama_obfuslm.tex
\begin{table*}[!t]
\centering
\begin{threeparttable}

 \caption{Attacking examples on two settings, obfuscated by ObfusLM~\cite{DBLP:conf/acl/LinYMZH0WLCD025}.}
\label{tab:atk_examples_llama_obfuslm}

\setlength{\tabcolsep}{2mm}
\renewcommand{\arraystretch}{1.2}
\begin{tabular}{@{}p{0.23\textwidth}|p{0.26\textwidth}|p{0.24\textwidth}|p{0.23\textwidth}@{}}
\toprule
 \multicolumn{4}{@{}l}{\textit{Setting: ObfusLM ($\epsilon=0.1$) + RoBERTa + SST-2}} \\
\midrule
\textbf{Original} & \textbf{Prompt Inv. (TBS~\cite{DBLP:conf/uss/Dong00C0Z25})} (Top-1: 0.164) & \textbf{InvBert~\cite{DBLP:journals/corr/abs-2109-10104}} (Top-1: 0.262) & \textbf{\NAME} (Top-1: 0.735)
\\
\midrule
\noindent\highlight{although laced with humor and a few fanciful touches , the film is a refreshingly serious look at young women .} &%
\noindent named \highlight{though} intertwined as quizz To's bunch crazensus moving,,,,,,,,,F\highlight{ilm} they's \highlight{refresh}ality grimL\highlight{ook}s as poorest ladies…. &%
\noindent as\highlight{though} tied as challenging to \highlight{a} bunch craziar moving ii. \highlight{film} i're impersonality grim \highlight{look}s as wealthiest ladies : &%
\noindent \highlight{though} combined \highlight{with humor and a few} funniful moving \highlight{, the} movie \highlight{is a refreshingly} grim \highlight{look at young women .}
\\
\midrule
 \multicolumn{4}{@{}l}{\textit{Setting: ObfusLM ($\epsilon=1.0$) + LLaMA3-8B + Medical QA}} \\
\midrule
\textbf{Original} & \textbf{Prompt Inv. (TBS~\cite{DBLP:conf/uss/Dong00C0Z25})}\newline{\footnotesize Top-1: 0.216, S-Top1: 0.205} & \textbf{InvBert~\cite{DBLP:journals/corr/abs-2109-10104}}\newline{\footnotesize Top-1: 0.457, S-Top1: 0.345} & \textbf{\NAME}\newline{\footnotesize Top-1: 0.804, S-Top1: 0.769}
\\ 
\midrule
\noindent\highlight{A} \sensitive{32-year-old woman} \highlight{comes to the emergency department because of a 3-hour history of severe nausea, vomiting, tremor, and anxiety. She recently started a new medication but does not remember its name. She has a history of} \sensitive{major depressive disorder} \highlight{treated with} \sensitive{fluoxetine}\highlight{. Her }\sensitive{temperature} \highlight{is} \sensitive{38.9 C (102.1 F)}\highlight{, }\sensitive{pulse} \highlight{is} \sensitive{132/min}\highlight{, }\sensitive{respirations} \highlight{are} \sensitive{22/min}\highlight{, and }\sensitive{blood pressure} \highlight{is} \sensitive{152/94 mm Hg}\highlight{. She is confused. Physical examination shows diaphoresis and an ataxic gait. }\sensitive{Patellar reflexes} \highlight{ are} \sensitive{4+ bilaterally}\highlight{.} &%
\noindent E's34année outdated \sensitive{woman} dominanceT\highlight{o} T\highlight{he emergency department} daß ofs das's VIII hours \highlight{History} ofs ***** \highlight{nausea}**, nausea**, tremblingARB*, AND Autism \%. elle Recent restarted das \highlight{new} drug \highlight{but} \highlight{does}** commemorate their \highlight{name} \%. elle exists das H\highlight{istory} ofs minor \red{depress}ion deficiency tratawith plaguedonavir \%. \highlight{Her} \sensitive{temperature}s mill's**\%.****289++. One cf), \_\sensitive{pulse} mill's130/\sensitive{min}**, respir affiliation \highlight{are}'s20/\highlight{min}**, and B\sensitive{lood}P\sensitive{ressure} mill's153()/88 inches K Rw\%. elle mill \highlight{confus}ing\%. \highlight{Physical examin}e \highlight{show} diaphphereis AND An Ataxvic(d Jill\%.Natural st\sensitive{ellar}.adv*** \highlight{are}'s*+ \sensitive{bilater} morally\%. &%
\noindent\highlight{A}934\sensitive{-year} detailed \sensitive{woman} dominates \highlight{to the emergency} engineer whereas of de-o VIII \highlight{hour}s \highlight{history of} potent \highlight{nausea}, nausea, \highlight{tremor}b, \highlight{and} autism\%. \highlight{She recently started} de \highlight{new} drug \highlight{but does not} realize her \highlight{name}\%. \highlight{She} got \highlight{a history of} minor \sensitive{depress}ion deficiency \highlight{treated with} diseaseoxetine\%. His \sensitive{temperature} mill s cresc\%.8 B\},290\%. One cf), \sensitive{pulse} mar d130 \sensitive{min}, \sensitive{respir}flation \highlight{are} in25e\sensitive{min}, \highlight{and} \sensitive{blood}-\sensitive{pressure} mill'154\%\%98 inches Bg+. \highlight{She} mill \highlight{confus}ing \% \highlight{Physical examination show}ed diaphferis \highlight{and an} ataxc nait \% Nat\sensitive{ellar} advives \highlight{are} in\sensitive{4+} bilter phys \% &%
\noindent\highlight{A} \sensitive{32-year-old woman} \highlight{comes to the emergency department because of a 3-hour history of} potential diarrhea, \highlight{vomiting}, andor, \highlight{and} depression. \highlight{She} has \highlight{started a new medication but does not remember} their \highlight{name.} \highlight{She has a history of} minor \sensitive{depressive disorder} \highlight{treated with} \sensitive{fluoxetine}\highlight{.} \highlight{Her} \sensitive{temperature} \highlight{is} 99°C8°C (289.7 ft)\highlight{, }\sensitive{pulse} \highlight{ is} \sensitive{1}10\sensitive{/min, respirations} \highlight{are }\sensitive{2}0\sensitive{/min}\highlight{, and }\sensitive{blood pressure} \highlight{ is} \sensitive{15}5\sensitive{/9}5 \sensitive{mm Hg. She is confused.} Her \highlight{examination shows} diapheresis, \highlight{an ataxic gait.} Her\sensitive{ellar} \sensitive{reflex}us \highlight{are} 5+ \sensitive{bilaterally}\highlight{.}
\\
\bottomrule
\end{tabular}
\begin{tablenotes}[flushleft]\footnotesize
\item[\dag] \highlight{Blue} text marks correctly recovered tokens; \sensitive{red} text marks sensitive tokens annotated via GPT-5.6. S-Top1 is the top-1 recovery rate restricted to sensitive tokens (see Section~\ref{sec:metrics}).
\item[\ddag] In the Prompt Inv.\ column, asterisks (\texttt{*}) replace non-ASCII characters produced by the inversion that cannot be properly rendered.
\end{tablenotes}
\end{threeparttable}
\end{table*}

%% file: sections/background.tex
\section{Background}

\subsection{Language Models}
Language models estimate the probability distribution over token sequences, typically by predicting missing or future tokens from their context. They are either bidirectional (e.g., masked language models like BERT) or autoregressive (e.g., GPT-style next-token predictors). Large language models (LLMs) are such models scaled to billions of parameters and trained on massive text corpora~\cite{Zhao2023Survey}. Below, we briefly review the Transformer architectures underlying both BERT-like models and autoregressive LLMs.


\noindent\textbf{Transformer backbone.} Modern language models are built on the Transformer architecture~\cite{VaswaniSPUJGKP17, Lin2021Survey}. Encoder-only models (e.g., BERT) use bidirectional self-attention for classification tasks, while decoder-only models (e.g., GPT, LLaMA, Qwen) use causal masking for autoregressive generation.

\noindent \textbf{LM-based downstream tasks.} BERT-style models are widely used for downstream classification tasks, in part because their parameter scales are moderate while still sufficient for many applications, including sentence-level classification (e.g., sentiment analysis) and token-level classification (e.g., named entity recognition). In contrast, autoregressive LLMs are typically fine-tuned to support generative tasks, where the model produces free-form text by iteratively predicting the next token.

\input{tables/obfus_tech}
\subsection{Obfuscation-Based Defenses}

To protect clients' prompt privacy, existing approaches can be broadly categorized into obfuscation-based defenses and cryptographic defenses. In this paper, we focus on obfuscation-based methods and defer a discussion of cryptographic techniques to Section~\ref{sec:discussion}. Below, we summarize several representative obfuscation techniques (also listed in Table~\ref{tab:obfus-tech}).

\noindent \textbf{Vocabulary permutation (VP).} Several defenses permute the vocabulary (and correspondingly the rows of the embedding matrix) so the server no longer knows which token each index represents. On its own, this offers limited protection, as permuted embeddings can often be re-identified by matching them to a public embedding space. Consequently, vocabulary permutation is usually paired with additional obfuscation mechanisms.

\noindent \textbf{Clustering/Mixing (CM).} Another strategy aims to confuse an adversary by mixing the embeddings of multiple tokens. Depending on how this mixing is implemented, the client may first cluster similar tokens and then use a cluster-level representation, or directly mix (e.g., average or linearly combine) embeddings across tokens or sentences to obscure the contribution of any single token.

\noindent \textbf{Token substitution (TS).} In token substitution, each token is randomly replaced by another token, where substitution probabilities are derived from a token similarity measure (e.g., in embedding space). This sampling is often implemented via the exponential mechanism from differential privacy~\cite{DBLP:conf/focs/McSherryT07} (usually a relaxed or informal variant), making similar tokens more likely substitutes.

\noindent \textbf{Noise adding (NA).} Noise can be added directly to token or sentence embeddings to achieve differential privacy, often termed \textit{embedding-level DP}. The client sends only the perturbed embeddings to the server.

\subsubsection{Attacks}
Embedding inversion attacks aim to recover the original tokens from obfuscated hidden states. Existing approaches can be broadly categorized into three classes based on their attack strategies.

\noindent \textbf{Hidden state comparison (KNN~\cite{DBLP:conf/cikm/QuKY0BN21}, EDNN~\cite{DBLP:conf/emnlp/LinZCHYLD24}).} The simplest attack strategy compares obfuscated hidden states against a public embedding space using distance metrics. 

\noindent \textbf{Direct prompt embedding optimization (ER, TBS~\cite{DBLP:conf/uss/Dong00C0Z25}, DEML~\cite{DBLP:conf/sp/0004ZWXYLZ25}).} These methods formulate token recovery as an optimization problem that minimizes the distance between the hidden states produced by candidate embeddings and the observed obfuscated states.

\noindent \textbf{Model-based inversion (InvBert~\cite{DBLP:journals/corr/abs-2109-10104}, MLC~\cite{DBLP:conf/ccs/SongR20}).} The most sophisticated approach trains an explicit inversion model to map obfuscated representations back to tokens. Our work advances this category by introducing a new inversion paradigm for highly accurate token recovery.

%% file: tables/obfus_tech.tex
\begin{table}[!t]
\small
\centering

\begin{threeparttable}
\caption{Defenses and their obfuscation techniques}
\label{tab:obfus-tech}

\begin{tabular}{@{}l|c|l|c}
\toprule
Defense & Tech. & 
Defense & Tech.
\\
\midrule
SentinelLMs$^\text{\textdagger}$~\cite{DBLP:conf/aaai/MishraLD24}    &  VP &
Datamix~\cite{DBLP:conf/eccv/LiuWGZH20}        &  CM
\\
Santext~\cite{DBLP:conf/acl/YueDWLSC21}        &  TS &
Custext~\cite{DBLP:conf/acl/ChenMWCN0C23}        &  TS
\\
TextObf.~\cite{DBLP:conf/acl/ZhouLMGWDZZH23} &  CM, NA &
TextMixer$^\text{\textdaggerdbl}$~\cite{DBLP:conf/emnlp/ZhouLMG0H23}      &  VP, CM
\\
ObfusLM~\cite{DBLP:conf/acl/LinYMZH0WLCD025}        &  VP, CM &
DPNR~\cite{DBLP:conf/emnlp/LyuHL20}           &  NA
\\
CAPE~\cite{DBLP:conf/emnlp/PlantGG21}          &  NA &
DP-Forward~\cite{DBLP:conf/ccs/DuYC0H023}    &  NA
\\

\bottomrule
\end{tabular}

\begin{tablenotes}[flushleft]
\footnotesize
\item[] \textdagger: VP is applied with a transformation named glide-reflection in SentinelLMs.
\item[] \revised{\textdaggerdbl: TextMixer relies on data multiplexing (MUX-PLMs) and is architecturally incompatible with the standard obfuscation pipeline evaluated in this work; we list it for completeness but do not include it in our experiments.}

\end{tablenotes}

\end{threeparttable}

\end{table}

%% file: sections/design.tex
\section{Problem Setting and Motivation}

\begin{figure*}[h]
\centering
\setlength{\abovecaptionskip}{10pt}
\setlength{\belowcaptionskip}{-5pt}


\includegraphics[width=0.85\textwidth]{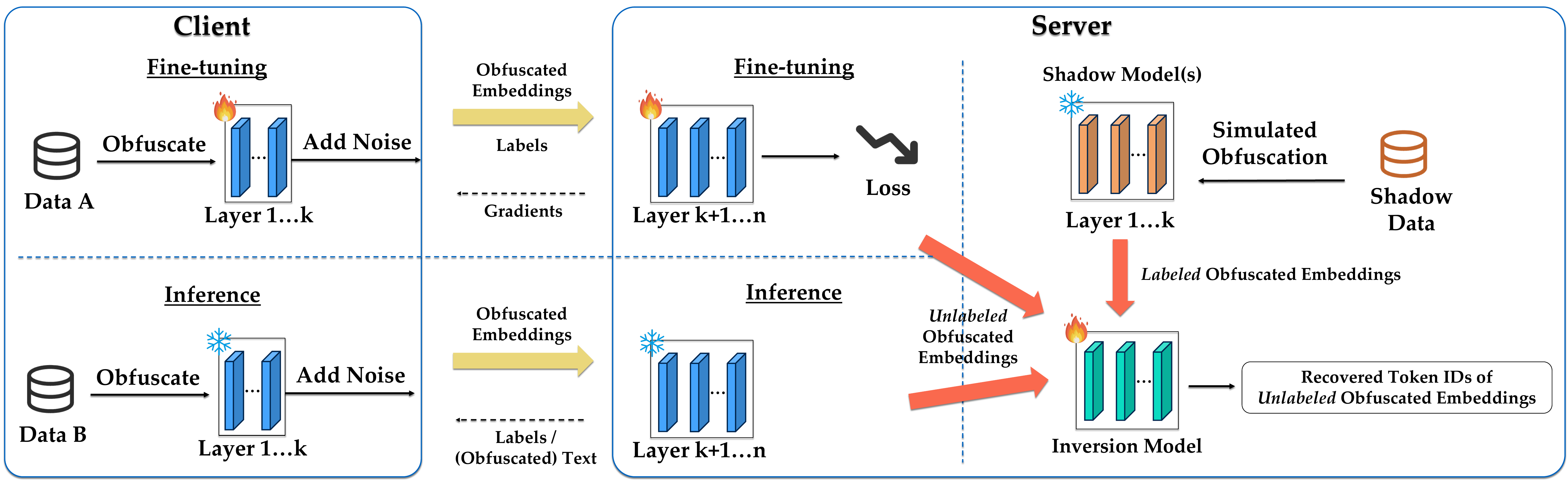}

\caption{Obfuscated LM Defense and Threat Model. Obfuscated embeddings (without original token IDs) are sent from client to server. The adversary simulates the obfuscation on shadow data to obtain labeled obfuscated embeddings, which are then used to train an inversion model in a supervised manner.}\label{fig:framework}
\end{figure*}

\subsection{System and Threat Model}
We provide an overview of existing obfuscated LM defense systems and the associated threat model in Figure~\ref{fig:framework}. A language model is partitioned into two components: a small prefix $F_c$ (layers 1 through $k$) deployed on the client side, and the remaining, much larger suffix $F_s$ deployed on the server side. There are typically two phases in which privacy must be preserved: (i) the fine-tuning phase and (ii) the inference phase. In both phases, only obfuscated embeddings are transmitted from client to server. \textbf{\textit{Notably, this also covers the basic setting with $k=0$, where only shuffled token IDs are transmitted~\cite{DBLP:conf/acl/LinYMZH0WLCD025, DBLP:conf/aaai/MishraLD24}; in this case, the client hosts no layer and incurs no computational overhead.}} The obfuscation procedure is identical across phases; the only difference is that, during fine-tuning, model parameters on both the client and server are updated via backpropagation. Our work is agnostic to how the task model is optimized and does not consider information leakage via gradient updates, which constitutes a separate attack surface that has been extensively studied~\cite{DBLP:conf/sp/MelisSCS19, DBLP:conf/ccs/SongR20}.

The client aims to protect its sensitive dataset or prompts from the server, which we treat as an adversary. \textit{This threat model captures the scenario where a client outsources model hosting to the cloud: the model itself need not be confidential, but the client's prompts (e.g., medical queries, financial records) must remain private.} The adversary observes all transmitted obfuscated embeddings, which are not labeled with their original token IDs. Moreover, the adversary is assumed to know the architecture of the task model and can train shadow models on a shadow dataset. The shadow dataset may be drawn from the same distribution as the client's data or, under a more constrained setting, from an unrelated public out-of-distribution corpus. Importantly, the shadow dataset need not be task-related; for the attack, it suffices that the adversary can generate \textit{labeled} obfuscated embeddings, which requires only text prompts, not ground-truth task labels. Because the adversary does not have access to all obfuscation information retained locally by the client, it can only approximate or simulate the obfuscation mechanism when generating labeled obfuscated embeddings. The adversary's goal is to recover the original token IDs corresponding to the observed unlabeled obfuscated embeddings.

\subsection{Problem Formulation}\label{sec:problem_formulation}
\revised{
\noindent\textbf{Security and utility goals.} Following prior works, the client's security goal is \emph{confidentiality of prompt content}: the server should not reconstruct original tokens from obfuscated embeddings. The utility goal is that obfuscated embeddings remain sufficiently informative for the downstream task. We adopt \emph{token-level recovery accuracy} (top-1) as the primary security metric: if the attacker recovers individual tokens, it can potentially extract sensitive entities, topics, or phrases. This is a strict, conservative measure—a defense that fails under this metric also fails under any weaker notion of confidentiality.
}

\noindent\textbf{Attack formulation.} A client holds a sensitive prompt $x = (x_1, x_2, \ldots, x_n)$ consisting of $n$ tokens. To protect prompt privacy, the client applies an obfuscation mechanism $\mathcal{O}(F_c, x)$ before transmitting them to the server.


Given an obfuscated hidden state $A = \mathcal{O}(F_c, x)$ received from the client, the server aims to reconstruct a prompt $x' = (x'_1, x'_2, \ldots, x'_n)$ such that $x' \approx x$. The fundamental challenge stems from the information loss introduced by obfuscation. Formally, the server seeks to solve:
\begin{equation}
\min_{x' \in \mathcal{V}^n} \left\| \mathcal{O}'(F_c, x') - A \right\|_2^2,
\label{eq:attack_objective}
\end{equation}
where $\mathcal{V}$ is the vocabulary set and $\mathcal{O}'$ represents the server's simulation of the obfuscation mechanism. However, this discrete optimization is \revised{computationally intractable} with a naive searching technique due to the combinatorial search space $|\mathcal{V}|^n$, which grows exponentially in the sequence length.

Prior work such as InvBert~\cite{DBLP:journals/corr/abs-2109-10104} relaxes this problem by training a supervised inversion model on labeled data, but suffers from distribution shift between shadow and target data. Direct embedding optimization methods (e.g., ER, TBS~\cite{DBLP:conf/uss/Dong00C0Z25}, DEML~\cite{DBLP:conf/sp/0004ZWXYLZ25}) optimize continuous embeddings but lack structural robustness to obfuscation noise.


\revised{
\noindent\textbf{Obfuscation mechanisms.} The defenses we evaluate instantiate $\mathcal{O}$ in four main ways (Table~\ref{tab:obfus-tech}):

\noindent(1) \emph{Vocabulary permutation (VP).} The client applies a secret bijection $\pi: \mathcal{V} \to \mathcal{V}$ to token IDs and correspondingly permutes the rows of the embedding matrix $E$: $E' = E_{\pi}$, where $E_{\pi}[i] = E[\pi(i)]$. The server observes permuted IDs (or their embeddings) but not $\pi$.

\noindent(2) \emph{Clustering/Mixing (CM).} Tokens are grouped into clusters $\mathcal{C} = \{C_1, \ldots, C_M\}$. Given a token $x_i$ belonging to cluster $C(x_i)$, its embedding is replaced by a fused representation, e.g., $h'_i = \mathrm{mix}\bigl(\{h_j : x_j \in C(x_i)\}\bigr)$, which may be a cluster centroid or a weighted combination of member embeddings.

\noindent(3) \emph{Token substitution (TS).} Each token $x_i$ is replaced by a substitute $x'_i \in \mathcal{V}$ sampled with probability
\[
\Pr[x'_i = v \mid x_i] \propto \exp\!\left(\frac{u(x_i, v)}{2\Delta u}\right),
\]
where $u(\cdot,\cdot)$ is a utility function (e.g., embedding distance) and $\Delta u$ is the sensitivity, under a (relaxed) exponential mechanism~\cite{DBLP:conf/focs/McSherryT07}.

\noindent(4) \emph{Noise adding (NA).} The client computes hidden states $h = F_c(x)$, normalizes them, and adds calibrated noise:
\begin{equation}\label{eq:na_defense}
    h' = \mathrm{normalize}(h) + e, \quad e \sim \mathrm{DP}(\epsilon),
\end{equation}
where $e$ is drawn from a DP mechanism with privacy budget $\epsilon$. Defenses differ in their normalization: DPNR uses min-max, CAPE uses $L_1$, DP-Forward uses Frobenius normalization; TextObfuscator omits normalization, leaving sensitivity unbounded.
}

\begin{figure*}[!htbp]
  \centering
  \subfloat[Original/Obfuscated Embeddings]{
	\fbox{\includegraphics[width=0.23\textwidth]{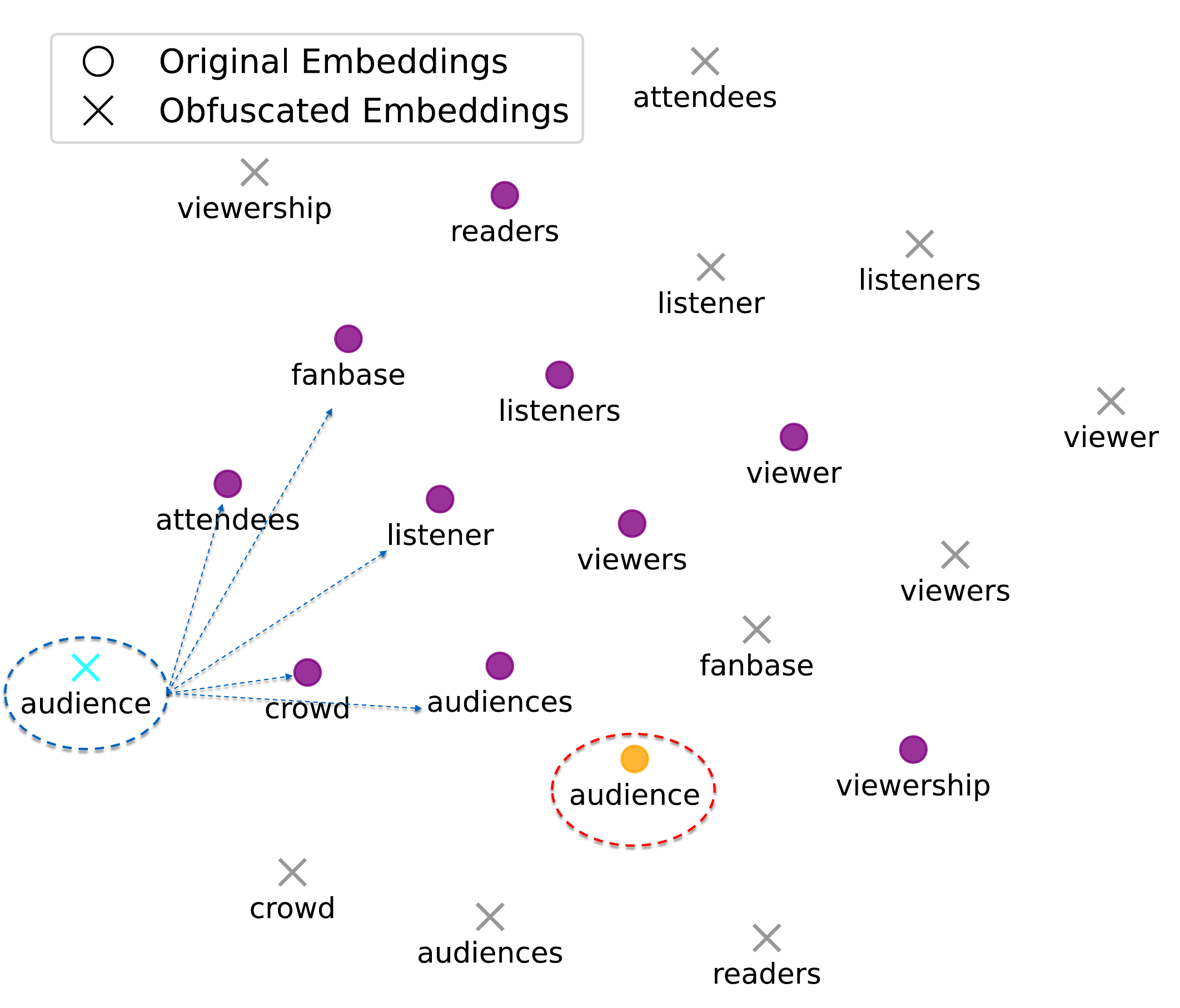}}
  }
  \subfloat[Hidden States (Original)]{
	\fbox{\includegraphics[width=0.23\textwidth]{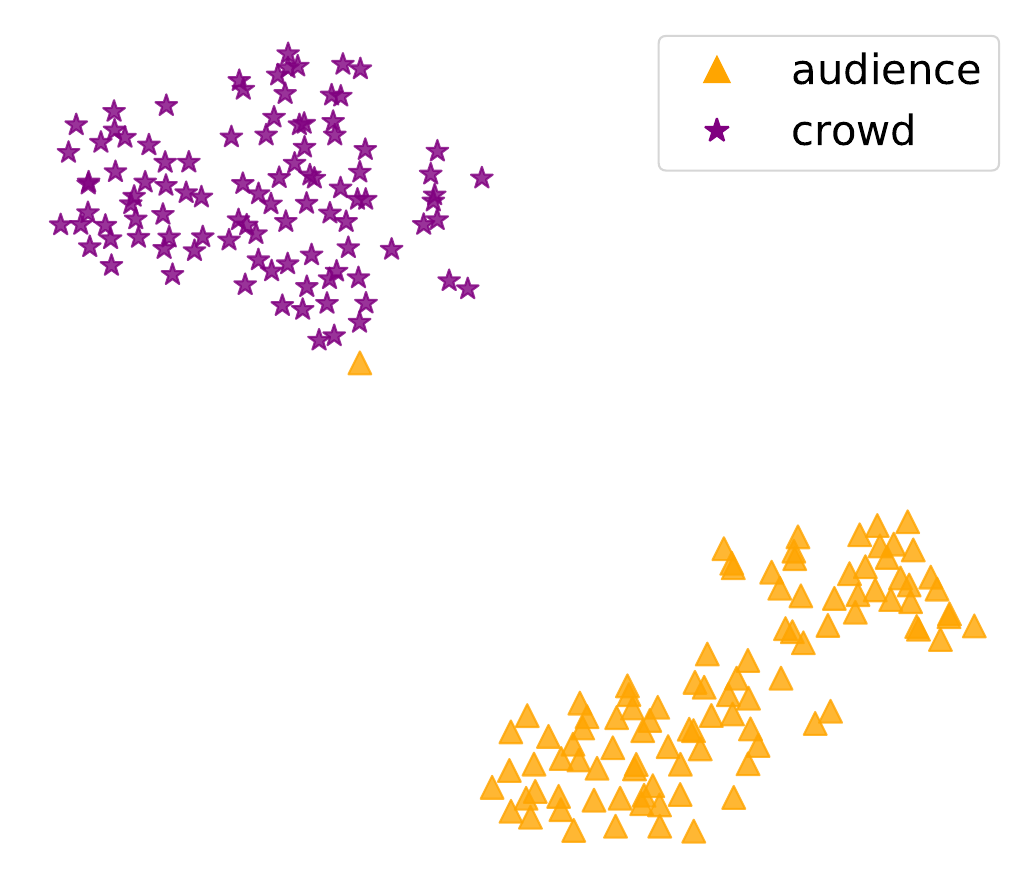}}
  } 
    \subfloat[Hidden States (Obfuscated)]{
	\fbox{\includegraphics[width=0.23\textwidth]{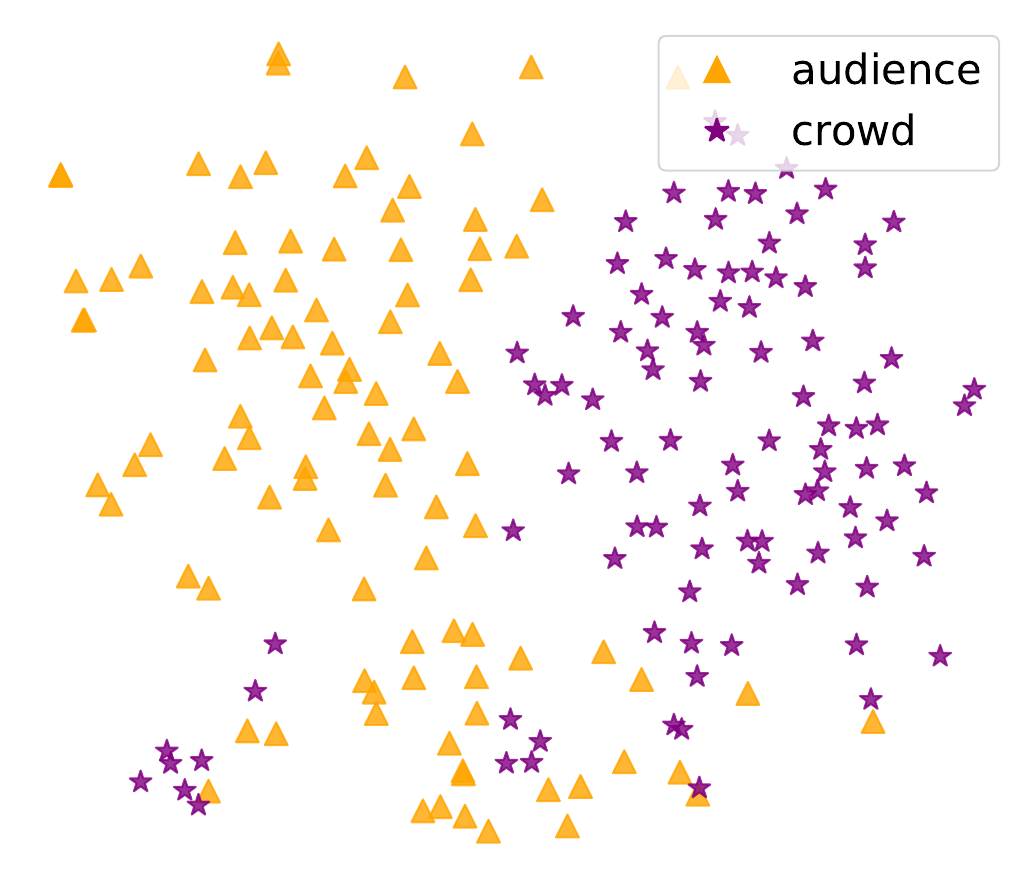}}
}
    \subfloat[InvBert Attack]{
    \fbox{\includegraphics[width=0.23\textwidth]{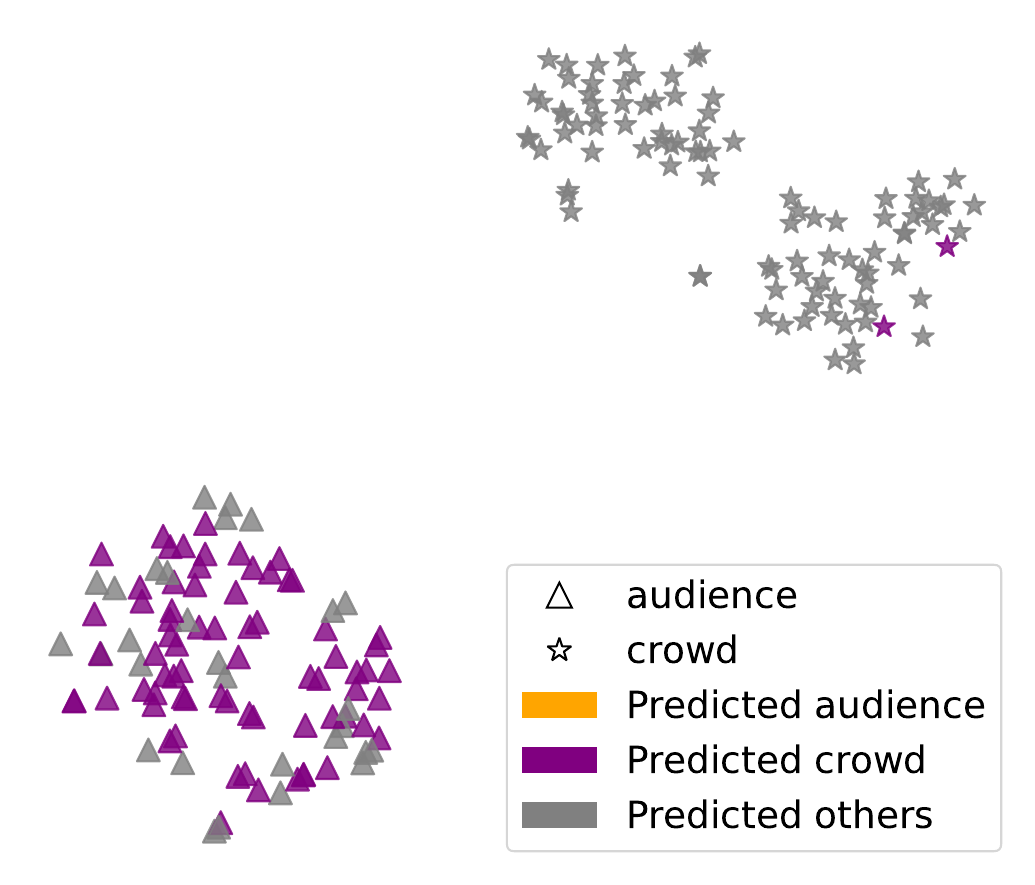}}
}
  \caption{
  (a) The original embedding of ``audience'' (\orange{orange $\bullet$}) is not among the top-5 nearest neighbors of its obfuscated embedding (\cyan{cyan $\boldsymbol{\times}$}). (b) The original hidden states for ``audience'' and ``crowd'' are clearly separable. (c) The obfuscated hidden states of ``audience'' and ``crowd'' are not clearly separable. (d) InvBert separates the obfuscated hidden states but misclassifies most samples. Each point represents the hidden state of one token occurrence (``audience'' or ``crowd'') in a different sentence.}\label{fig:neighbors}
\end{figure*}

\subsection{Motivating the Design}

We use the token ``audience'' as an example to illustrate limitations of current defenses and attacks. Figure~\ref{fig:neighbors}-(a) shows that ObfusLM's obfuscation moves embeddings away from their original positions, making nearest-neighbor attacks like KNN ineffective. The original ``audience'' embedding does not even appear among the top-5 nearest neighbors of the obfuscated one. However, the relative distances among tokens are largely preserved to maintain downstream task utility. This creates a false sense of security: while individual embeddings are perturbed, the semantic structure in hidden states remains exploitable. Figures~\ref{fig:neighbors}-(b,c) compare BERT's last-layer hidden states for ``audience'' and its neighbor ``crowd'' before and after obfuscation. The defense blurs but does not eliminate the semantic separation, suggesting that a more sophisticated attack could recover this structure.

Figure~\ref{fig:neighbors}-(d) shows InvBert's attack results. Though it learns to separate obfuscated hidden states, it fails to assign correct token IDs in most cases due to insufficient exploitation of semantic structure and overfitting to shadow data. This motivates our semi-supervised approach that leverages both labeled shadow data and unlabeled target embeddings to discover invariant semantic features.

%% file: sections/attack.tex
\section{\tNAME}\label{sec:method}

We highlight several main issues that lead to suboptimal results in the InvBert attack. First, the supervised learning approach places less focus on discovering potentially unseen semantic structure among noisy embeddings, and targets at an easier task of linking  embeddings to their labeled token IDs. Second, noisy embeddings of prompts used in the online service (i.e., the test phase) preserve such semantic information, but they can not be used in training InvBert because their real token IDs are unknown. Finally, for some defenses, the inversion training requires simulating a shadow defense model that might behave differently from a real defense model,
causing a severe problem of distribution shift.

Inspired by the discovery described above, we propose a novel design, \NAME, based on semi-supervised learning to leverage rich semantic information from both labeled and unlabeled data. 

\subsection{Semi-Supervised Inversion Attack}
Let $f_\theta: \fin \rightarrow \fout$ denote the inversion model parameterized by $\theta$, where $L$ is the input (prompt) length, $H$ is the embedding size, and $V=|\mathcal{V}|$ is the vocabulary size. Given an input sequence $x \in \fin$, we denote by $f_\theta(x)_{i, v}$ the predicted probability of token $v$ at position $i$. We draw inspiration from a classical semi-supervised learning algorithm \textit{FixMatch}~\cite{DBLP:conf/nips/SohnBCZZRCKL20}. 
For labeled data, the inversion model is trained in a standard supervised manner with cross-entropy loss $\mathcal{L}_s$:
\begin{equation}\label{eq:sup}
    \mathcal{L}_s = - \frac{1}{L}\sum_{i=1}^{L} \log f_\theta(x)_{i, y_i},
\end{equation}
where $y_i \in \{1, \dots, V\}$ is the ground truth token ID on position $i$. FixMatch improves performance on unlabeled data via \textit{pseudo-labeling}. In summary, it produces a weakly augmented view and a strongly augmented view for each unlabeled sample. High-confidence predictions on the weak views are used as pseudo labels to supervise the training of strong views, termed as the \textit{unsupervised consistency regularization loss} $\mathcal{L}_u$ (also a cross-entropy loss). FixMatch was designed for image classification, where supervised training often yields reliable high-confidence pseudo-labels. However, we have a more challenging task to predict correct tokens in a large vocabulary (typically tens of thousands) given potentially heavily perturbed embeddings. A direct application of FixMatch yields little benefit, so we adopt a tailored design. We introduce three key enhancements here, and defer additional optimizations to later sections.


\begin{figure}[!t]
  \centering
  \subfloat[Inversion on TextObfuscator]{
	\fbox{\includegraphics[width=0.45\linewidth]{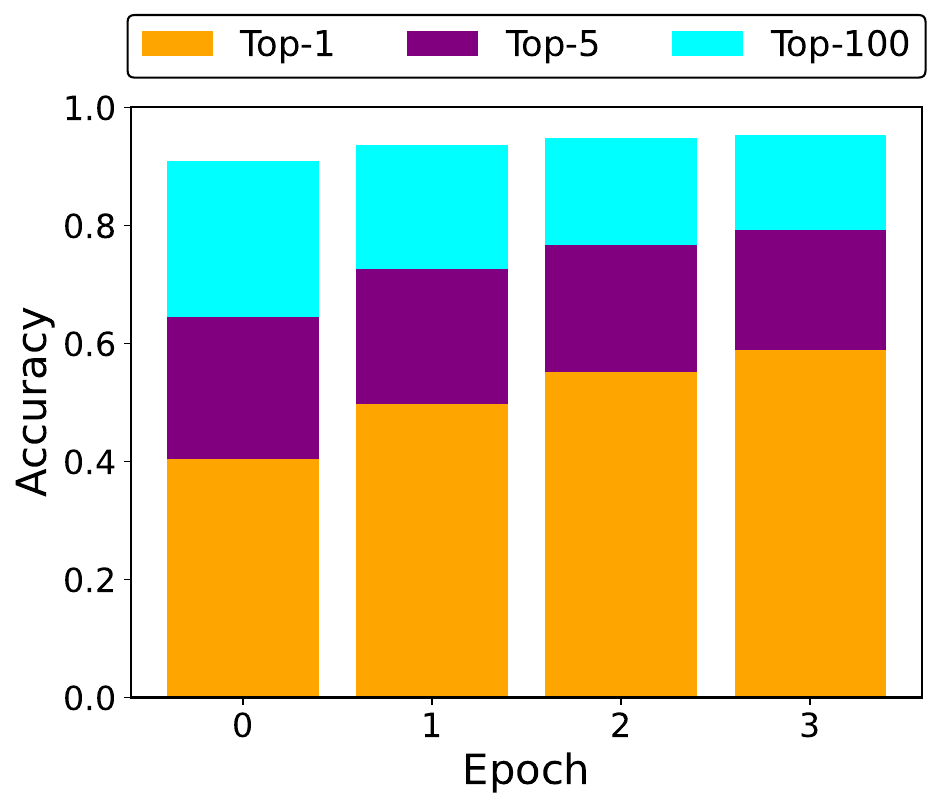}}
  }
  \subfloat[Inversion on ObfusLM]{
	\fbox{\includegraphics[width=0.45\linewidth]{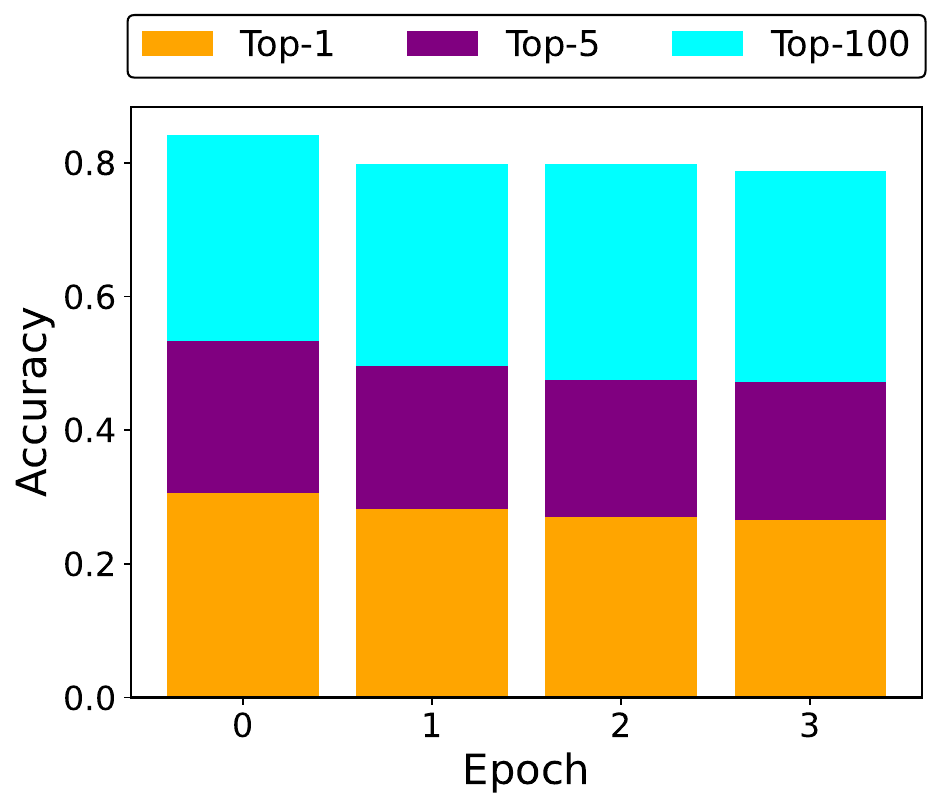}}
  }
  \caption{Test accuracy of supervised attacks. Attacker models achieve high top-100 matching accuracy overall. For ObfusLM, the attack overfits to the shadow distribution, resulting in decreasing test accuracy as training progresses.}
  \label{fig:epoch_acc}
\end{figure}

\subsubsection{Soft Labeling}
While analyzing the performance of supervised inversion learning, we find that it achieves relatively good results after the first few epochs if we consider a loose metric such as top 100 accuracy for each token (see Figure~\ref{fig:epoch_acc}). Such an inversion model might not be useful for an attacker to recover a prompt, but it provides a good starting point of teacher signals.

Instead of using highly inaccurate hard pseudo labels, we propose to use \textit{soft labeling}. Given a weakly augmented view $x_w$ and a strongly augmented view $x_s$ of the unlabeled input $x$, a direct approach is to minimize KL-divergence loss on the two views to enforce consistency regularization: 
\begin{align*}
    \mathcal{L}_u
    & = \frac{1}{L}\sum_{i=1}^{L} D_{\kl}(f_\theta(x_s)_{i} || f_\theta(x_w)_{i}) \\
    & = \frac{1}{L}\sum_{i=1}^{L}\sum_{v=1}^{V}  f_\theta(x_s)_{i, v} \log \frac{f_\theta(x_s)_{i, v}}{f_\theta(x_w)_{i, v}},
\end{align*}

Given the observation in Figure~\ref{fig:epoch_acc}, we further constrain the loss in the range of top-$k$ highest probabilities, which both optimizes computation resources and minimizes the influence of long-tail low-confidence predictions. For each sample $x \in \fin$, let $\mathcal{I}_{x, i}^{(k)}$ denote the indices of the top-$k$ largest entries in $f_\theta(x)_{i}$, and let $f_\theta(x)_{i, \mathcal{I}_{x, i}^{(k)}}$ denote these top-$k$ probabilities at position $i$. The top-$k$ KL divergence loss is defined as:
\begin{equation}\label{eq:unsup-topk}
    \loss_u^{(k)}
    = \frac{1}{L}\sum_{i=1}^{L} D_{\kl}(f_\theta(x_s)_{i, \mathcal{I}_{x_w, i}^{(k)}} || f_\theta(x_w)_{i, \mathcal{I}_{x_w, i}^{(k)}})
\end{equation}
Note that the indices $\mathcal{I}_{x_w, i}^{(k)}$ for both probability distributions come from the prediction on the view $x_w$. The final loss is:
\begin{equation}\label{eq:joint}
    \loss = \loss_s + \lambda \cdot \loss_u^{(k)},
\end{equation}
where $\lambda$ is the strength parameter of unsupervised loss.

\subsubsection{Masked Augmented Views}

Most defenses cause some damage to semantic information due to the perturbation added to the inputs, such as clustering, noise adding, and vocabulary shuffling. A critical step to improve inversion performance is to restore such semantic information on the perturbed embeddings. We borrow the token masking idea from pretraining masked language models (e.g., BERT), but apply it to embeddings instead of raw tokens. Essentially, for input $x \in \fin$, we randomly sample a percentage of positions and replace their embeddings with the $\mask$ embedding. 
\begin{align}
    & r = (r_1, \dots,r_L)\in \mathbb{R}^L, r_i \sim \mathcal{U}(0, 1) \quad \text{i.i.d for } i=1,\dots, L \\
    & x_{i, :} \leftarrow e_{\mask} \quad \text{if} \quad r_i < p_m,
\end{align}
where $e_{\mask} \in \RR^H$ is the embedding of the $\mask$ token (randomly initialized if $\mask$ token does not exist in the tokenizer), and $p_m$ is a hyperparameter for masking probability.


We obtain strongly and weakly augmented views $x_w$ and $x_s$ with masking under different parameters $p_m$. Equation~\ref{eq:unsup-topk} can thus be instantiated with $x_s$ and $x_w$. In addition, we apply the weakly augmented views to the supervised component, which helps reduce overfitting to the labeled data distribution. In this case, $x$ should be replaced with $x_w$ in Equation~\ref{eq:sup}.

\subsubsection{Smooth Update with EMA Teacher}
In the above unsupervised loss (Equation~\ref{eq:unsup-topk}), we are forwarding $x_w$ and $x_s$ to the same up-to-date model $f_\theta$, basically instructing the model to always learn from itself. However, directly fitting to the up-to-date model tends to be unstable and even cause model collapse, since the target distribution can change abruptly from step to step, causing optimization to chase a rapidly moving target instead of learning consistent representations~\cite{tarvainen_mean_2017, DBLP:conf/nips/GrillSATRBDPGAP20}. 
We instead use an exponential moving average (EMA) teacher~\cite{tarvainen_mean_2017}, which provides a more stable training target by aggregating the student's parameters over time, thereby smoothing out the high-variance updates from recent mini-batches.  EMA teachers have been widely adopted in semi-supervised learning and self-distillation.

Let $f_{\theta_t}$ denote a teacher inversion model and $f_{\theta_s}$ denote a student inversion model. The unsupervised KL divergence loss is further refined to be:
\begin{equation}\label{eq:final_sup}
    \loss_u^{(k)}
    = \frac{1}{L}\sum_{i=1}^{L} D_{\kl}(f_{\theta_s}(x_s)_{i, \mathcal{I}_{x_w, i}^{(k)}} || f_{\theta_t}(x_w)_{i, \mathcal{I}_{x_w, i}^{(k)}})
\end{equation}
The teacher model is updated using an exponential moving average:
\begin{equation}
    \theta_t \leftarrow \alpha \theta_t + (1-\alpha)\theta_s,
\end{equation}
where $\alpha \in [0, 1)$ is the smoothing coefficient controlling how much the teacher follows the student, e.g., $\alpha=0.998$. The supervised training component always updates the student model $f_{\theta_s}$.

\subsection{Mixed Training: Mutual Reinforcement}

Unfortunately, the joint optimization (Equation~\ref{eq:joint}) does not work well in our setting. At the beginning of training, the model has very low accuracy in predicting token IDs, and the KL divergence loss therefore provides misleading signals that push the model in many incorrect directions. Simply disabling this loss in the first few epochs slightly alleviates the problem, but the model still saturates at a suboptimal performance level towards the end of training. This phenomenon has been systematically studied in multi-task learning~\cite{DBLP:conf/nips/YuK0LHF20}, where different task losses are optimized jointly (on training data with potentially different distribution). Joint optimization might suffer from gradient conflicts between objectives $\mathcal{L}_s$ and $\mathcal{L}_u^{(k)}$, which can slow down training and drive the model to sub-optimal compromise solutions. Rather than tuning a fixed $\lambda$, we adopt an alternating schedule that avoids per-step interference.

To better optimize both objectives, we follow a simple yet effective \textit{mixed training} strategy. The core idea is to alternate the two objectives and let the model focus on one objective at a time according to a predefined schedule. This also enables the two components to reinforce each other with newly learned knowledge. We introduce an unsupervised ratio $\tau_i \in [0,1]$ for epoch $i$, which denotes the fraction of optimization steps (mini-batches) allocated to unsupervised training. We compute the ratio $\tau_i$ using a linear schedule:
\begin{align}
\tau_i \leftarrow \tau_0 + \frac{i}{n-1} \cdot (\tau_{n-1} - \tau_0), \label{eq:epoch_ratio}
\end{align}
where $n$ is the total number of epochs. $\tau_0$ and $\tau_{n-1}$ can be adjusted to reflect a global training budget for each component. Given that each epoch is trained for $E$ steps, in epoch $i$, we first optimize the supervised objective for $(1-\tau_i) \cdot E$ steps, and then switch to optimizing the unsupervised loss for $\tau_i \cdot E$ steps. Our approach can be interpreted as a curriculum over objectives~\cite{DBLP:conf/icml/BengioLCW09, DBLP:journals/corr/abs-2101-10382}. We provide an extended discussion of this design in Appendix~\ref{sec:app_mixed}, including its motivation as an alternating optimization scheme and a budget interpretation.

\begin{algorithm}[!t]
  \caption{\tNAME}
  \label{alg:method}
  \KwIn{Inversion model $f_\theta$; Real target model $d_{\theta_{\mathrm{real}}}$;
        Shadow target model $d_{\theta_{\mathrm{shadow}}}$; Target layer $l$;
        Victim's prompt dataset $D_{\mathrm{real}}$; Attacker's prompt dataset $D_{\mathrm{shadow}}$;
        Fine-tuning epochs $n$; Epoch optimization steps $E$; Learning rate $\eta$;
        KL divergence size $k$; Smoothing coefficient $\alpha$}
  \KwOut{Student model parameter $\theta_s$}

  $\theta_t \leftarrow \theta$\;
  $\theta_s \leftarrow \theta$\;

  \For{$i \gets 1$ \KwTo $n$}{
    Compute epoch-level ratio $\tau_i$ with Eq.~\ref{eq:epoch_ratio}\;

    \tcp{Supervised training}
    \For{$j \gets 1$ \KwTo $(1-\tau_i)\cdot E$}{
      Sample a $\mathsf{prompt}$ from $D_{\mathrm{shadow}}$\;
      \tcp{Extract a labeled tuple $x\in \fin, y \in \mathcal{V}^L$}
      $x, y \leftarrow \mathrm{extract}(d_{\theta_{\mathrm{shadow}}}, \mathsf{prompt}, l)$\;
      \tcp{Generate weakly augmented view}
      $x_w \leftarrow \mathrm{augment}(x)$\;
      $\Delta \theta_s \leftarrow \nabla_{\theta_s}
         \loss_s(f_{\theta_s}(x_w), y)$\tcp*{Eq.~\ref{eq:sup}}
      $\theta_s \leftarrow \theta_s - \eta \Delta \theta_s$\;
      $\theta_t \leftarrow \mathrm{sync}(\theta_t, \theta_s, \alpha)$\;
    }

    \tcp{Unsupervised training}
    \For{$j \gets (1-\tau_i)\cdot E + 1$ \KwTo $E$}{
      Sample a $\mathsf{prompt}$ from $D_{\mathrm{real}}$\;
      \tcp{Extract an unlabeled $x\in \fin$}
      $x \leftarrow \mathrm{extract}(d_{\theta_{\mathrm{real}}}, \mathsf{prompt}, l)$\;
      \tcp{Generate weakly and strongly augmented views}
      $x_w, x_s \leftarrow \mathrm{augment}(x)$\;
      $\Delta \theta_s \leftarrow \nabla_{\theta_s}
        \loss_u^{(k)}(f_{\theta_s}(x_s), f_{\theta_t}(x_w))$\tcp*{Eq.~\ref{eq:final_sup}}
      $\theta_s \leftarrow \theta_s - \eta \Delta \theta_s$\;
      $\theta_t \leftarrow \mathrm{sync}(\theta_t, \theta_s, \alpha)$\;
    }
  }
\end{algorithm}

\noindent\textbf{Finalized algorithm.} We outline \NAME~training in Algorithm~\ref{alg:method} and highlight additional details below. Suppose the obfuscation-based system requires clients to upload layer-$l$ embeddings (including the word embedding layer). To obtain embeddings $x \in \fin$ (and optionally $y\in \mathcal{V}^L$ for labeled data), a text prompt is fed into the target model (real or shadow), and the layer-$l$ hidden states are extracted. For clarity, we omit standard training details such as mini-batching, learning-rate scheduling, and optimization hyperparameters. Further enhancements to this basic algorithm are introduced in the next section.

\section{Optimizations}
In this section, we introduce several optimizations and extensions in order to adapt \NAME~to different categories of defenses. 

\subsection{Multi-Shadow Attack}
\begin{figure}[!t]
  \centering
  \subfloat[InvBert Attack]{
	\fbox{\includegraphics[width=0.45\linewidth]{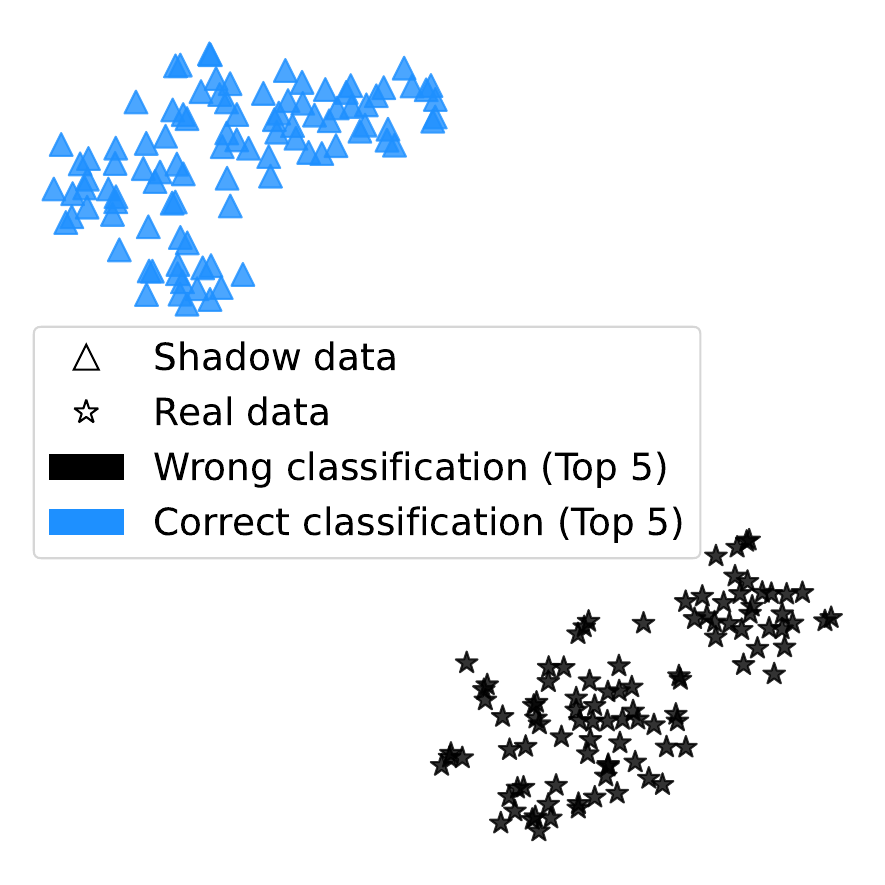}}
  }
  \subfloat[\NAME~Attack]{
	\fbox{\includegraphics[width=0.45\linewidth]{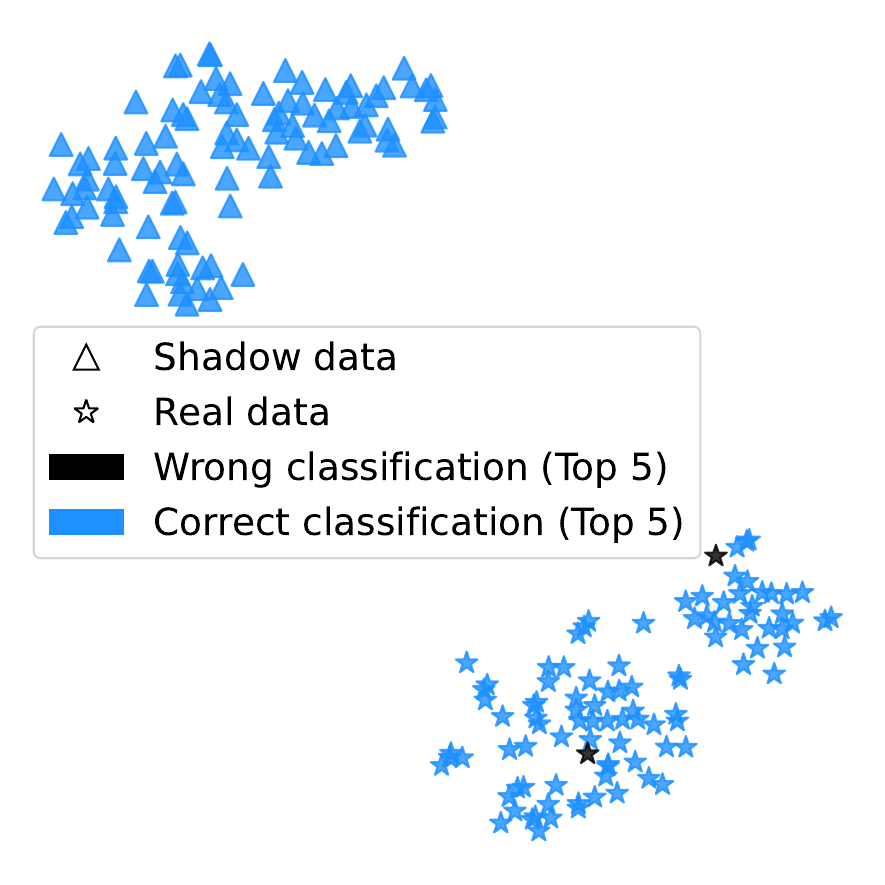}}
  }
  \caption{Distribution of shadow and real embeddings for the token ``audience'' and the corresponding classification outcomes. (a) InvBert attains high accuracy on shadow embeddings but consistently misclassifies real embeddings due to distribution shift. (b) \NAME~correctly classifies most real embeddings. Both shadow and real embeddings are derived from data not seen during training.}
  \label{fig:real_vs_shadow}
\end{figure}

In order to train the supervised component, the attacker needs access to obfuscated token embeddings for input sentences. In other words, the attacker must know the \textit{white-box} obfuscation model so that it can generate obfuscated embeddings for chosen tokens. However, this white-box assumption is unrealistic in settings where the client holds a secret, such as a private vocabulary permutation~\cite{DBLP:conf/acl/LinYMZH0WLCD025}. Similar to InvBert, \NAME~uses shadow models to remove this assumption. A shadow model employs its own randomness to generate obfuscated embeddings for supervised training, with the hope that these \textit{shadow} obfuscated embeddings follow a similar distribution to the \textit{real} obfuscated embeddings. In practice, however, the inversion model does not transfer well to real obfuscated embeddings because of the distribution shift (see Figure~\ref{fig:real_vs_shadow}).

During supervised training, the inversion model tends to learn a tight decision boundary that encloses the shadow embeddings of each token, inevitably excluding many real embeddings that are distributed differently. To encourage the inversion model to capture deeper, invariant features (e.g., semantic information) that remain stable despite differences in randomness between a shadow model and the real model, \NAME~leverages multiple shadow models instantiated with different randomness to substantially improve inversion performance. The attacker's capabilities and assumptions remain unchanged; the only additional requirement is extra resources (e.g., computation and memory) to run multiple shadow models.


\subsection{Renormalization and Denoising}
Differential privacy provides a strong defense with formal guarantees and has been incorporated into several obfuscated language model designs~\cite{DBLP:conf/acl/ZhouLMGWDZZH23, DBLP:conf/emnlp/LyuHL20, DBLP:conf/emnlp/PlantGG21, DBLP:conf/ccs/DuYC0H023}. While such schemes are theoretically hard to fully compromise, we show how \NAME~can be further strengthened to improve its attack performance against DP-based defenses. 

\revised{In embedding-level DP-based defenses, the obfuscated representation follows Eq.~\eqref{eq:na_defense}: for a plain input $x$, the client transmits $x^* = \text{normalize}(x) + e$, where $e$ is noise drawn from a DP mechanism and ``normalize'' bounds the sensitivity of outputs.}

\textbf{Renormalization.} Existing obfuscation defenses apply different normalization schemes to hidden representations, such as min-max normalization~\cite{DBLP:conf/emnlp/LyuHL20}, $L_1$ normalization~\cite{DBLP:conf/emnlp/PlantGG21}, and Frobenius normalization~\cite{DBLP:conf/ccs/DuYC0H023}. These schemes rescale the input $x$ to different magnitudes and inject noise with scales determined by the privacy parameter $\epsilon$. As a result, the effective noise level and representation scale can vary substantially across defenses, making it difficult to select augmentation parameters that produce meaningful alternative views of a given embedding $x^*$. To address this, we first renormalize $x^*$ using layer normalization before applying any augmentation. This preprocessing step maps embeddings to a more uniform scale, allowing us to use a single augmentation parameter that performs robustly across a wide range of defenses. A further advantage is that this renormalization improves compatibility with diverse model architectures, including many autoregressive generators (e.g., LLaMA, Qwen) whose hidden states follow a \emph{pre-norm} design, i.e., normalization is applied at the layer input rather than the output. As we demonstrate in our experiments, this renormalization step has a substantial impact on the effectiveness of \NAME.

\textbf{Denoising.} The main difficulty of inverting the noisy embeddings lies in the low \textit{signal-to-noise ratio} (SNR). 
After normalizing the embeddings, the standard deviation of noises is usually non-negligible with appropriately chosen $\epsilon$ parameters. However, embeddings are highly structural vectors where each dimension contains nontrivial signals that enable differentiation of tokens, whereas noises in DP are zero-mean isotropic. Since an attacker can generate unlimited number of ``clean'' embeddings and their corresponding ``noisy'' embeddings with shadow models, it can first use classical denoising solutions to lower the energy of noise before feeding into \NAME.

\textit{Principal component analysis} (PCA) turns out to be a perfect fit in this case and is often used as a standard linear denoising solution. Let $K$ denote the number of projected dimensions and $P \in \RR^{H \times K}$ denote the PCA reduction matrix. We estimate $P$ based on a large volume of plain shadow embeddings. After PCA transformation, embeddings are projected to principal components where standard deviation is largest, while noises are effectively averaged and thus have reduced standard deviation. 

\begin{lemma}
    Suppose $g \in \RR^H$ is a noise vector where each dimension is independent with mean 0 and variance $\sigma^2$, i.e., $g \sim (0, \sigma^2I_H)$. Let $z=PP^Tg$. Then $\frac{\EE\|z\|^2}{\EE\|g\|^2} = \frac{K}{H}$.
\end{lemma}

We give a brief proof for this lemma in Appendix~\ref{sec:app_pca}. As for an embedding $h \in \RR^H$, it follows from the literature~\cite{bishop:2006:PRML} that this ratio $\frac{\EE\|PP^Th\|^2}{\EE\|h\|^2}$ before and after the projection is roughly $\frac{\sum_{i=1}^K\lambda_i}{\sum_{i=1}^H\lambda_i}$, where $\{\lambda_i\}$ are eigenvalues. The ratio could be configured to be close to 1 with modest $K$ because PCA concentrates most of the energy in the eigen directions with top-$K$ largest eigen values. Therefore, in order to smooth out noise in $x^*$, we apply the transformation $PP^Tx^*$, which essentially reduces the energy of noise while keeping the energy of embeddings mostly unchanged.  Note that this does not constitute an attack on differential privacy, but suggests a more stringent choice of the privacy parameter $\epsilon$ in the scenario of obfuscating language models.


\subsection{Inversion Models}
To minimize the effort of distribution adaption and leverage pretrained knowledge, \NAME~instantiates the backbone of the inversion model as that of the defended target model. In this work, we target transformer-based models and verify the effectiveness of our approach across both classification and generation scenarios.

For classification tasks that generally rely on smaller models (e.g., BERT), we use full-parameter fine tuning (Full-FT). For generation models with billions of parameters, we use LoRA~\cite{HuSWALWWC22} to enable Parameter-Efficient Fine-Tuning (PEFT) due to limited computational and memory resources. \NAME~remains effective under both fine-tuning paradigms.

%% file: sections/evaluation.tex
\section{Evaluation}

\subsection{Setup}
\subsubsection{Prototype} 
We implemented a prototype of \NAME\ in PyTorch~\cite{PaszkeGMLBCKLGA19} following Algorithm~\ref{alg:method}, and ran all experiments on four NVIDIA H800 GPUs with 80GB of memory each. In addition, we integrate the prototype in a modularized library that enables extending various defenses and attacks efficiently. 
Detailed hyperparameters (augmentation parameters, learning rates, batch sizes, epochs, etc.) are provided in Appendix~\ref{sec:app_implementation}.

\subsubsection{Models}
We evaluate \NAME\ on four pretrained models covering both classification and generation tasks.

\noindent\textbf{BERT-Base~\cite{DBLP:conf/naacl/DevlinCLT19}:} A bidirectional Transformer encoder pretrained with masked LM and next-sentence prediction. We use BERT-Base for classification. It has a 30K vocabulary, 768-dimensional embeddings, and 12 Transformer layers.

\noindent\textbf{RoBERTa-Base~\cite{DBLP:journals/corr/abs-1907-11692}:} A BERT-style encoder with the same architecture as BERT but a different pretraining scheme and a larger vocabulary of size 50K.

\noindent\textbf{LLaMA3-8B-Instruct~\cite{grattafiori2024llama}:} An autoregressive LLM based on the LLaMA3-8B architecture, used for generation experiments. It has a 128K vocabulary, 4096-dimensional embeddings, and 32 layers.

\noindent\textbf{Qwen3.5-27B~\cite{qwen35-27b}:} An autoregressive LLM built on the Qwen3.5 architecture, optimized for complex reasoning, code generation, and multi-turn dialogue. It has a 248K vocabulary, 5120-dimensional embeddings, and 64 Transformer layers.




\subsubsection{Datasets}\label{sec:dataset}
We evaluate \NAME~on five publicly available datasets spanning classification and generation tasks: SST-2~\cite{DBLP:conf/emnlp/SocherPWCMNP13} (binary sentiment classification), CoNLL-2003~\cite{DBLP:conf/conll/SangM03} (token-level named entity recognition), AG News~\cite{DBLP:conf/nips/ZhangZL15} (four-class topic classification), Medical Meadow MedQA~\cite{jin2020disease, medalpaca-medqa} (medical licensing exam QA), and Enron Emails~\cite{corbt-enron} (corporate email generation). For NER we report entity-level F1 using $\mathsf{seqeval}$~\cite{seqeval}. For all datasets, we follow prior work~\cite{DBLP:conf/acl/LinYMZH0WLCD025, DBLP:conf/acl/ZhouLMGWDZZH23, DBLP:conf/emnlp/ZhouLMG0H23} and report performance on a validation split whenever one is provided; otherwise we evaluate on the standard test split or a held-out 20\% partition. Full dataset descriptions, construction details (e.g., how we adapt Enron Emails for generation), and statistics are provided in Appendix~\ref{sec:app_dataset}.

\subsubsection{Defenses and Attacks}\label{sec:baseline}
We conduct a comprehensive evaluation of nine obfuscation-based language model defenses that cover a broad range of techniques: Datamix~\cite{DBLP:conf/eccv/LiuWGZH20}, SentinelLMs~\cite{DBLP:conf/aaai/MishraLD24}, Santext~\cite{DBLP:conf/acl/YueDWLSC21}, Custext~\cite{DBLP:conf/acl/ChenMWCN0C23}, TextObfuscator~\cite{DBLP:conf/acl/ZhouLMGWDZZH23}, ObfusLM~\cite{DBLP:conf/acl/LinYMZH0WLCD025}, DPNR~\cite{DBLP:conf/emnlp/LyuHL20}, CAPE~\cite{DBLP:conf/emnlp/PlantGG21}, and DP-Forward~\cite{DBLP:conf/ccs/DuYC0H023}. With the exception of Datamix and SentinelLMs---which rely solely on heuristic mechanisms such as sample mixing or geometric transformations (e.g., glide reflection)---these schemes incorporate some form of differential privacy. We organize our analysis in two stages. First, we show that several defenses become insecure under our attack even when configured with standard, commonly recommended parameters. We then provide a more detailed investigation of defenses that offer formal differential privacy guarantees.

As attack baselines, we include KNN~\cite{DBLP:conf/cikm/QuKY0BN21}, InvBert~\cite{DBLP:journals/corr/abs-2109-10104}, and three advanced prompt-inversion attacks ER, TBS~\cite{DBLP:conf/uss/Dong00C0Z25}, DEML~\cite{DBLP:conf/sp/0004ZWXYLZ25}. KNN is a highly efficient, non-parametric attack that directly compares obfuscated embeddings against a public embedding space and recovers tokens by nearest-neighbor search. InvBert, in contrast, trains a supervised inversion model to map obfuscated representations back to the underlying tokens. ER, TBS, and DEML directly optimize the prompts (in the embedding space) that result in observed hidden states. We do not include the MLC attack~\cite{DBLP:conf/ccs/SongR20} in our comparison because it shares a similar attack strategy with InvBert but reports a different metric—whether a sentence contains a given token—which is not directly comparable to the ordered token-level recovery metrics used in this work. For empirical results on InvBert and MLC under related settings, we refer readers to prior studies~\cite{DBLP:conf/acl/ZhouLMGWDZZH23, DBLP:conf/emnlp/ZhouLMG0H23}.

\subsubsection{Metrics}\label{sec:metrics}
\revised{
Top-$k$ measures the fraction of tokens for which the correct word appears among the attacker's top-$k$ predictions; we report top-1 as the strictest metric. We also report ROUGE-L and BLEU for sequence-level semantic similarity. Top-1 is the primary privacy metric, whereas ROUGE-L/BLEU serve as secondary indicators of semantic fidelity.

\noindent\textbf{Sensitive Top-1.} To further demonstrate the severity and effectiveness of the attack on sensitive content, and to validate the use of the standard top-1 metric, we introduce \emph{Sensitive Top-1}. We use GPT-5.6 to annotate sensitive tokens in the evaluation data (e.g., medical terms, named entities, numerical values; the tagging prompt is provided in Appendix~\ref{sec:app_sensitive_tagging}), and compute the top-1 recovery rate restricted to these annotated sensitive tokens. This metric is applied only to the Medical QA and Email datasets, which contain sensitive information. Our results (Table~\ref{tab:generation}) show that Sensitive Top-1 closely tracks the standard top-1 metric, confirming that the attack recovers sensitive and non-sensitive tokens at comparable rates.
}

\revised{
\subsubsection{Attack Cost and Evaluation Protocol}
For classification, training \NAME\ takes 2--5 hours on a single H800 GPU (25 epochs, 1 shadow) or 7--10 hours (50 epochs, 7 shadows for ObfusLM). For generation on LLaMA3-8B and Qwen3.5-27B with LoRA, training takes 24--48 hours on 4 H800 GPUs. InvBert uses the same backbone with a single shadow and 25 epochs.

\revised{\noindent\textbf{Transductive setting.} Our main experiments follow the transductive setting, where the adversary maximizes ASR by leveraging all available unlabeled target embeddings---consistent with prior prompt-inversion works~\cite{DBLP:conf/uss/Dong00C0Z25, DBLP:conf/sp/0004ZWXYLZ25}. We also provide ablation results in Section~\ref{sec:ablation_volume} where the transductive assumption is removed, showing that \NAME\ remains effective when transferring to held-out prompts.}
}

\input{tables/classification}

\subsection{Small Models for Classification Tasks}
\revised{
For the classification tasks, we report attack performance on all defenses in Table~\ref{tab:classification} (binary) and Table~\ref{tab:classification-multi} (multi-class, Appendix). Each row is a defense; bold indicates the best attack.
} To facilitate fair comparison, we set $\epsilon=3$ for all $\epsilon$-DP-based schemes. ObfusLM adopts a non-standard $(k, \epsilon)$-DP notion; we also evaluate $\epsilon=0.1$ as recommended in its original paper.

\subsubsection{Binary classification}
For binary classification (Table~\ref{tab:classification}), we present results on both BERT- and RoBERTa-based models. Under the recommended privacy settings, many defenses already incur a noticeable degradation in task accuracy. SentinelLMs is a notable outlier: it provides very weak protection and is essentially broken even by simple attacks, with KNN already achieving an average top-1 ASR above 0.91, and both InvBert and \NAME\ attaining essentially perfect recovery ($\ge 0.998$). \revised{\NAME~outperforms prior attacks on most defenses, with the magnitude of improvement varying by defense type.} The most striking gains occur on ObfusLM with $\epsilon=0.1$, where \NAME~reaches a top-1 recovery accuracy of 0.735, compared to the previous best of 0.262 by InvBert, effectively undermining the level of protection claimed by ObfusLM. The main limitation of InvBert on ObfusLM is its tendency to overfit to the shadow data distribution. In contrast, \NAME~mitigates this issue by exploiting noisy augmented views and multiple shadow models during training. TextObfuscator omits embedding normalization, leaving the sensitivity effectively unbounded; consequently, both InvBert and \NAME~achieve high ASR, with \NAME's advantage being more pronounced on RoBERTa than on BERT. We attribute this to differences in hidden state magnitudes across model architectures; a detailed analysis is provided in Appendix~\ref{sec:app_hidden_noise}. We give some concrete attack examples in Appendix~\ref{sec:app_atk_example}.

The improvements on Santext and Custext are relatively smaller. They apply token-level DP by randomly replacing tokens with other vocabulary items; consequently, \NAME~can only correct these corrupted positions by leveraging sentence-level semantic context, a capability also exploited by InvBert when instantiated with a language-model backbone. Nevertheless, our use of augmented views maintains an advantage over InvBert. Results are broadly similar between BERT and RoBERTa, despite RoBERTa's substantially larger vocabulary, suggesting that \NAME~is robust to moderate architectural and tokenizer differences.

The other three attack baselines (ER, TBS, and DEML) exhibit similar attack performance across all defenses, and in most cases underperform compared to both InvBert and \NAME. Given their comparable effectiveness, we include only TBS or DEML in subsequent experiments, and focus on InvBert and \NAME~to conserve computational resources and concentrate on the most insightful results. 

\subsubsection{Multi-class classification}
We extend the analysis to multi-class classification (AG News and CoNLL-2003 NER) in Appendix~\ref{sec:classification-multi}. \NAME\ attains the highest ASR against all defenses: e.g., on ObfusLM ($\epsilon=0.1$), it reaches an average ASR of 0.818 on AG News (vs.\ 0.267 for InvBert), and 0.644 on CoNLL-2003 (vs.\ 0.358), demonstrating stable generalization across heterogeneous obfuscation mechanisms.

\begin{figure*}[!t]
  \centering
  \subfloat[TextObfuscator SST-2]{
	\includegraphics[width=0.24\textwidth]{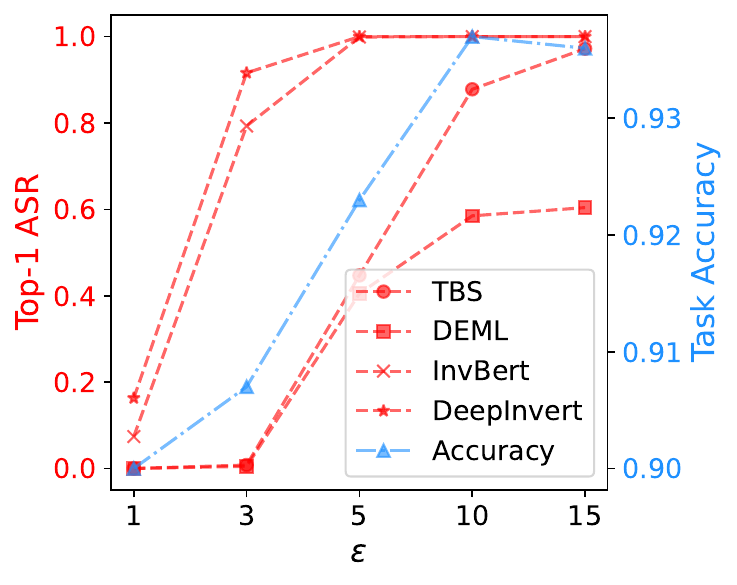}
  }
  \subfloat[DPNR SST-2]{
	\includegraphics[width=0.24\textwidth]{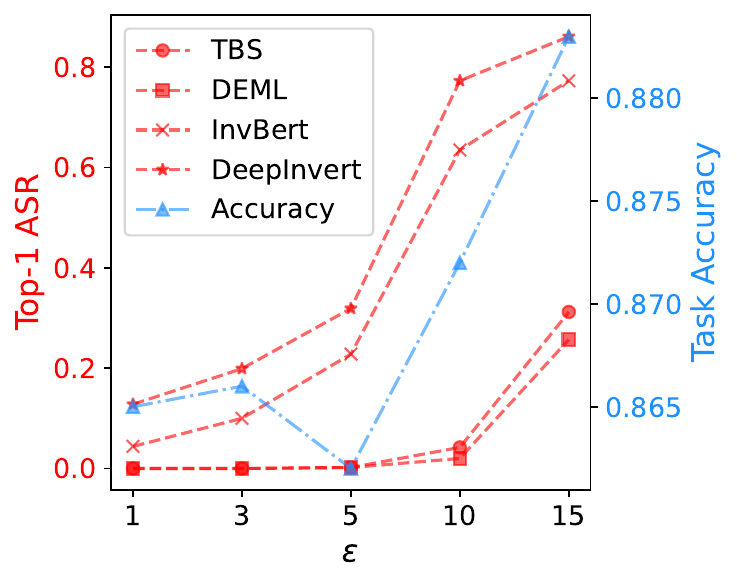}
  }
    \subfloat[CAPE SST-2]{
	\includegraphics[width=0.24\textwidth]{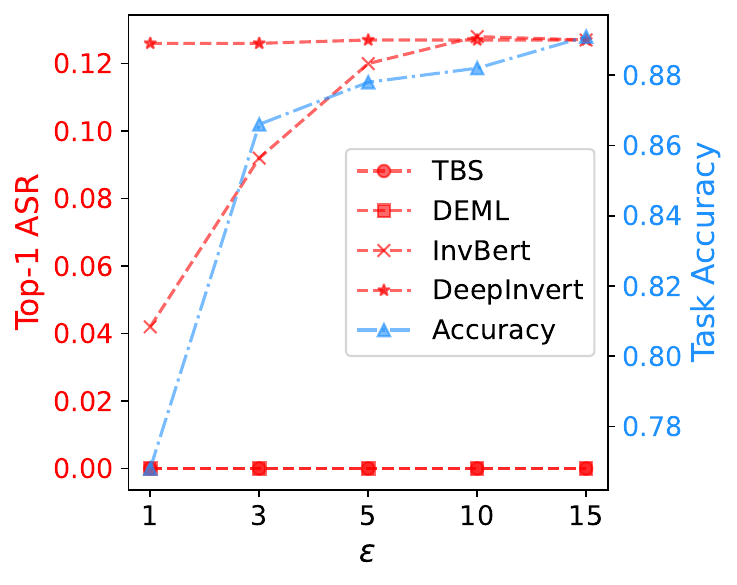}
  }
      \subfloat[DP-Forward SST-2]{
	\includegraphics[width=0.24\textwidth]{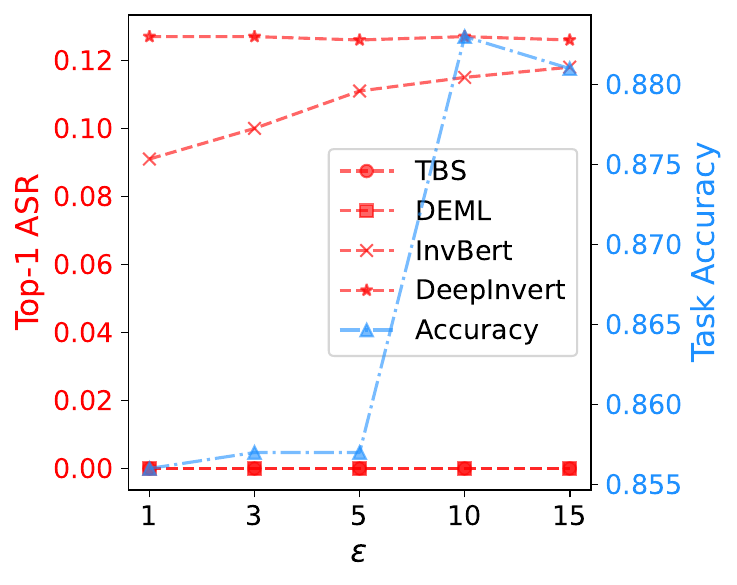}
  }
  \\

  \subfloat[TextObfuscator CoNLL]{
 \includegraphics[width=0.24\textwidth]{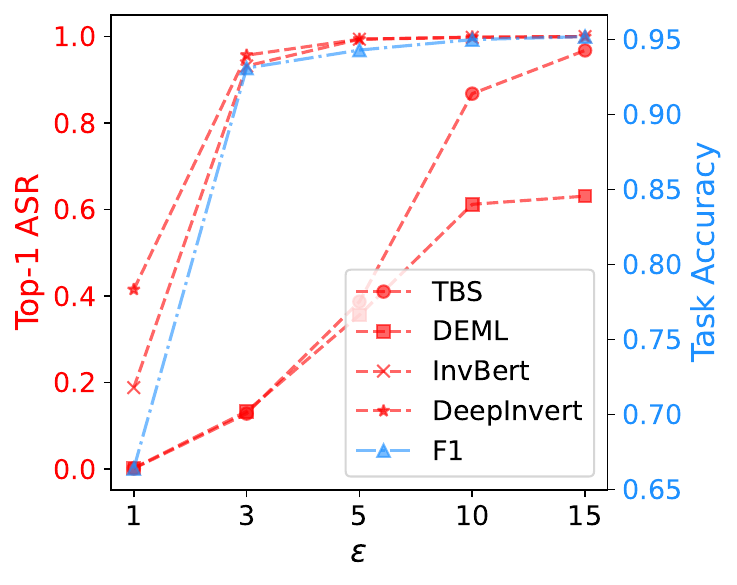}
  }
   \subfloat[DPNR CoNLL]{
	 \includegraphics[width=0.24\textwidth]{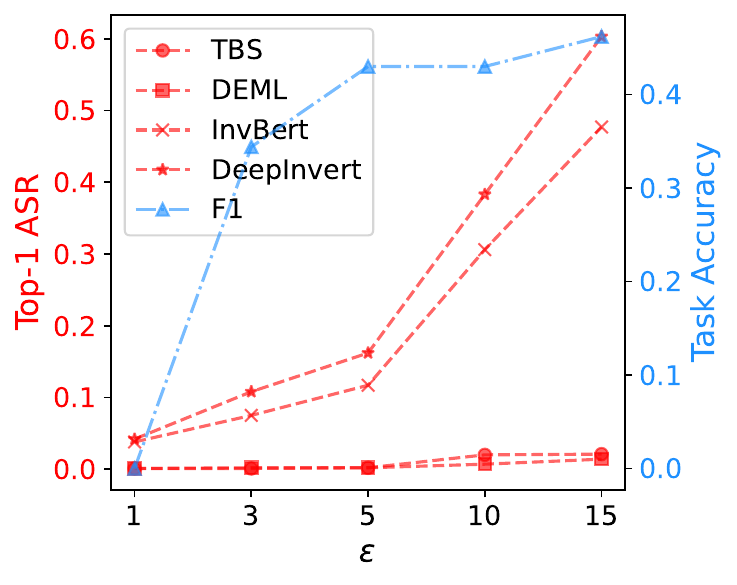}
   }
      \subfloat[CAPE CoNLL]{
	 \includegraphics[width=0.24\textwidth]{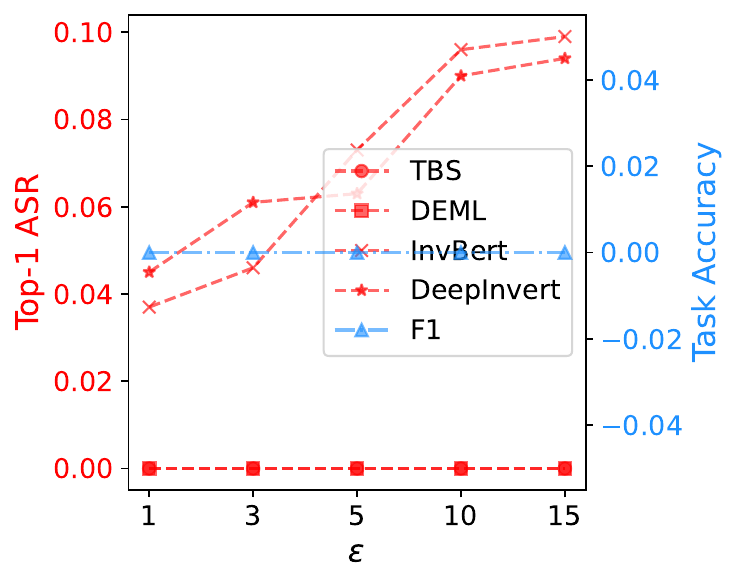}
   }
         \subfloat[DP-Forward CoNLL]{
	 \includegraphics[width=0.24\textwidth]{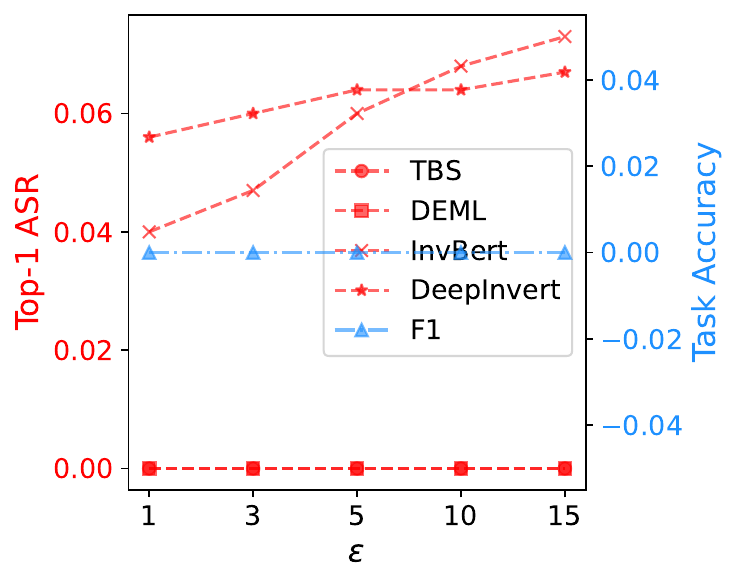}
   }

  \caption{Task and attack performance on embedding-level DP based schemes. Attack performance is aligned with the left Y axis, whereas task performance (accuracy or F1) is aligned with the right one.}\label{fig:dp}
\end{figure*}

 \subsubsection{Embedding-Level DP-based Defenses}\label{sec:eval_dp}
\revised{
To further understand the strengths and limitations of \NAME, we analyze its performance against embedding-level DP defenses, which constitute the strongest class of obfuscation-based protections among existing schemes. TextObfuscator, DPNR, CAPE, and DP-Forward fall into this category, but they differ significantly in the strength of their DP guarantees. TextObfuscator omits normalization entirely and therefore lacks any formal DP guarantee, as the sensitivity of the embedding mechanism is unbounded; its high ASR under \NAME\ thus reflects the defense's own miscalibration rather than a breakthrough by the attack. DPNR applies min-max normalization and bounds sensitivity at the coordinate level rather than for the full embedding vector, resulting in weaker protection than vector-level guarantees. For vector-level sensitivity, CAPE uses
$L_1$ normalization, while DP-Forward employs Frobenius (i.e.,
$L_2$-matrix) normalization; both provide formally stronger guarantees but at the cost of severe utility degradation.
} We summarize results on SST-2 and CoNLL-2003 in Figure~\ref{fig:dp}, varying $\epsilon$ from 1 to 15, corresponding to ``strong'' to ``moderate'' privacy regimes as characterized in prior works~\cite{DBLP:conf/acl/MeehanMC22, DBLP:conf/ccs/DuYC0H023}.

TextObfuscator becomes clearly insecure for $\epsilon \ge 3$, where \NAME\ attains an ASR above 0.9 on both SST-2 and CoNLL-2003. DPNR also fails to provide meaningful protection once $\epsilon$ reaches 10-15, especially on SST-2, where ASR grows substantially. CAPE and DP-Forward are comparatively more robust under the same $\epsilon$ values, consistent with their formal, vector-level DP guarantees. However, this robustness comes at the cost of severe utility loss. On the token-level CoNLL-2003 task, both CAPE and DP-Forward yield an F1 score of 0, effectively equivalent to random guessing. Thus, under their recommended $\epsilon$ settings, these schemes are not viable for token-level tasks. Achieving non-trivial task performance would require substantially larger $\epsilon$, at which point their effective protection degrades to a level comparable to DPNR. Note that when the DP noise is sufficiently large that InvBert's ASR becomes very low, \NAME\ also becomes unstable and may not exhibit a clear advantage, since its teacher model is no longer reliable.

Taken together, these experiments indicate that, with current embedding-level DP designs, the recommended $\epsilon$ regimes either leave \NAME\ with high recovery accuracy or collapse downstream task performance to near-random. This tension is even more pronounced in the generation setting, which is inherently a more challenging token-level prediction problem.

\input{tables/generation}

\subsection{Large Models for Generation Tasks}
We evaluate attacks on generation tasks using four representative defenses: ObfusLM ($\epsilon=1$), Santext ($\epsilon=3$), DPNR ($\epsilon=15$), and DP-Forward ($\epsilon=15$). We focus on these schemes for two reasons. First, several defenses are fundamentally incompatible with generation tasks: TextObfuscator and Datamix rely on output-space mixing or perturbation mechanisms that cannot be applied to autoregressive token generation. Second, some defenses such as SentinelLMs and Custext already proved highly vulnerable to all attacks in our classification experiments (with \NAME~achieving near-perfect recovery). Our experiments include Llama3-8B and Qwen3.5-27B, both of which have very large vocabularies.

In Table~\ref{tab:generation}, with the recommended $\epsilon=1$, ObfusLM is nearly completely compromised by \NAME\ that achieves a top-1 token recovery rate of 0.852 on Qwen3.5-27B in the medical QA dataset. Santext and DP-Forward were not originally designed or evaluated for generation tasks, and both exhibit low task utility in this setting (Rouge1 $\leq 0.54$ for Santext and $\approx 0$ for DP-Forward). For Santext, \NAME\ still consistently outperforms all baselines, though the margin over InvBert is modest, mirroring the pattern observed in our classification experiments: token-level replacement limits the additional gains from our unsupervised objective. DP-Forward remains robust at the recommended $\epsilon=15$ parameter, but at the cost of completely collapsing task utility. The inversion results on DPNR $\epsilon=15$ differ across models, due to its lack of theoretical guarantees for bounding sensitivity, leading to unstable relative noise magnitude compared to a model's hidden states. We provide a quantitative analysis in Appendix~\ref{sec:app_hidden_noise} explaining why the same defense can lead to substantially different inversion performance across different model architectures.

\revised{For the Medical QA and Email datasets, we additionally report Sensitive Top-1 (S-Top1 in Table~\ref{tab:generation}), which measures recovery restricted to sensitive tokens (e.g., medical terms, named entities, numerical values). Across all defenses and models, Sensitive Top-1 closely tracks the standard top-1 metric, confirming that \NAME\ recovers sensitive and non-sensitive tokens at comparable rates—there is no systematic gap that would suggest sensitive content is better protected.}

Given the similarity in attack effectiveness between our classification and generation results, we expect that other obfuscation defenses would exhibit comparable vulnerabilities if adapted to the generation setting without substantial redesign or strengthening of their schemes.

\subsection{Ablation Study}
In this section, we examine how the key components of \NAME\ affect its performance. Unless otherwise stated, experiments are conducted on SST-2 with a RoBERTa backbone.

\input{tables/ablation_basic}

\subsubsection{Effect of soft labeling and augmented views}
\revised{
We study the impact of soft labeling and masked augmented views in Table~\ref{tab:ablation-basic}, using ObfusLM ($\epsilon=0.1$) and DPNR ($\epsilon=10$) as representative defenses. The first row (neither masking nor soft labels) corresponds to a direct application of FixMatch to embedding inversion—i.e., a generic semi-supervised baseline without domain-specific adaptations. This configuration also serves as an approximate upper bound for InvBert with 7 shadow models, since it shares the same multi-shadow setup and training budget but additionally leverages unlabeled target embeddings via the FixMatch consistency objective. Our enhancements substantially outperform this baseline.
}

For ObfusLM, adding masked augmented views is crucial for the effectiveness of the unsupervised objective: it increases top-1 recovery accuracy from 0.377 to 0.702. Soft labeling further improves performance from 0.702 to 0.735 by reducing the risk of discarding correct labels when the inversion model is still imperfect. For DPNR, the gains are more modest, as embedding-level DP already injects substantial noise into the representations, effectively acting as a form of augmentation on the clean embeddings. Even so, combining soft labels and augmented views yields an additional two percentage points of attack accuracy.

\begin{figure}[!t]
  \centering
\subfloat[Effect of Shadows \& Strategy]{
\fbox{\includegraphics[width=0.46\linewidth]{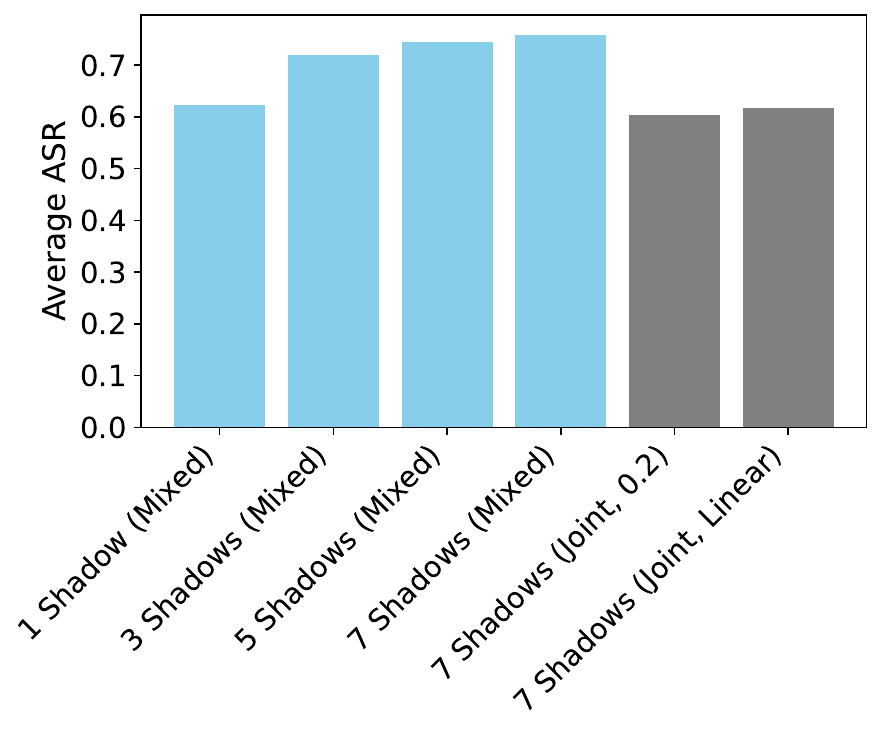}}
}
\subfloat[Effect of Target Layer]{
\fbox{\includegraphics[width=0.46\linewidth]{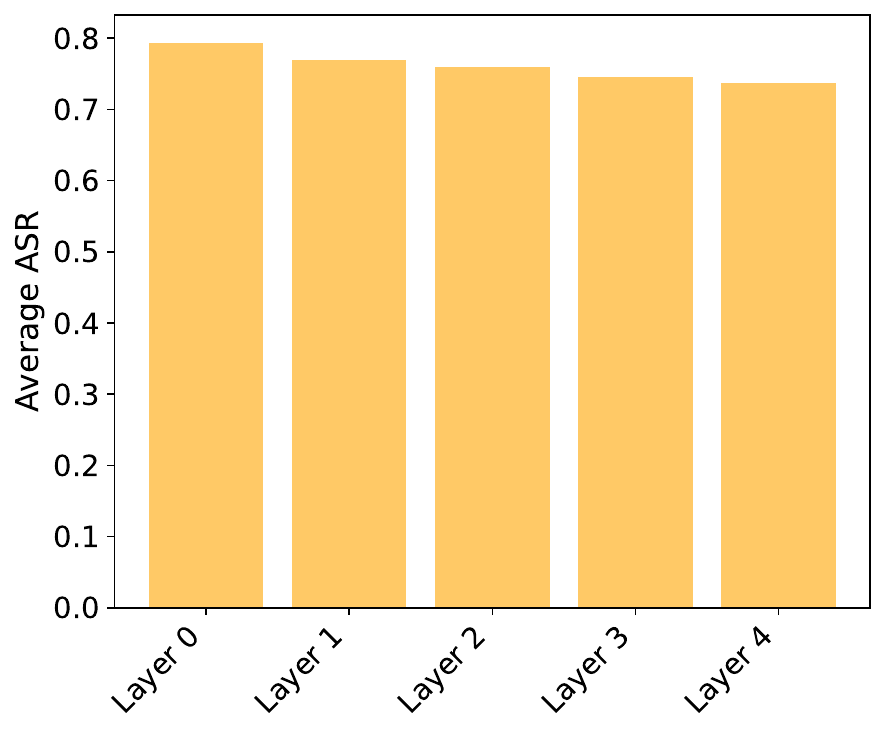}}
}
  \caption{(a) ASR increases with the number of shadow models, and the ``mixed'' strategy outperforms the ``joint'' strategy. (b) ASR slightly decreases when attacking deeper layers.}
  \label{fig:ablation_shadow_target}
\end{figure}

\subsubsection{Effect of shadow number and training strategy}
The multi-shadow technique is applied only to ObfusLM in our experiments, because its shadow and target embedding distributions differ substantially due to defense-side permutations and clustering. We therefore investigate how the number of shadow models affects attack performance. We also evaluate a joint optimization strategy (Eq.~\ref{eq:joint}), where $\lambda$ is either fixed at 0.2 or linearly increased from 0 to 1. Results for attacking ObfusLM ($\epsilon=0.1$) are shown in Figure~\ref{fig:ablation_shadow_target}-(a).

Using a single shadow model already yields an average ASR above 0.6, clearly outperforming InvBert (Table~\ref{tab:classification}), thanks to our semi-supervised components. Note that InvBert also uses a single shadow model and 25 epochs (same as the single-shadow \NAME\ configuration), so the comparison at $N_{\mathrm{shadow}}=1$ is compute-matched; the additional gains from more shadow models represent the added value of the multi-shadow technique. As the number of shadows increases, attack performance continues to improve, though with diminishing marginal returns. In the main experiments, we use seven shadows, which offers a good trade-off between effectiveness and computational cost.

For a fixed number of seven shadows, the joint optimization strategy is less effective than our mixed training design. We attribute this to more efficient use of the optimization budget in the mixed scheme. Nonetheless, we do not rule out that a more sophisticated scheduling policy for $\lambda$ could close this gap; exploring such schedulers is left for future work.

\subsubsection{Effect of target layer}
\revised{
From an architectural perspective (Figure~\ref{fig:framework}), the client typically retains a small number of initial layers, and obfuscates the hidden states produced at a chosen ``target'' layer. Our main experiments use target layer 3, where the client keeps layers $\{1,2, 3\}$. This is a common choice in prior work~\cite{DBLP:conf/acl/LinYMZH0WLCD025, DBLP:conf/acl/ZhouLMGWDZZH23} and reflects a practical trade-off: the client must retain enough layers to perform obfuscation, but offloads most computation to the server. We note that none of the evaluated defenses prescribes a specific target layer; the choice is left to the deploying client.
} In this section, we examine how attack performance varies with the target layer.

As shown in Figure~\ref{fig:ablation_shadow_target}-(b), obfuscating a deeper layer slightly weakens \NAME's effectiveness. This is expected: hidden states from later layers are more specialized for the downstream task, and thus carry less of the fine-grained, task-agnostic token semantics that inversion attacks exploit. Nevertheless, the ASR remains high even when the target layer is shifted to layer 5. We do not consider layers beyond 5, as doing so would undermine the intended setting in which the client offloads most of the computation to the cloud server.

\input{tables/ablation_denoise}
\subsubsection{Effect of renomalization and denoising}
Renormalization and denoising play an important role in enhancing \NAME's performance. We apply both operations to defenses that add embedding-level DP noise. In this section we use DPNR as a representative example, varying $\epsilon$. For denoising, we perform PCA on the obfuscated embeddings and project them to a reduced dimension $K=384$ (from the original 768). When the victim model does not normalize hidden states (as in some ObfusLM configurations), we also apply renormalization; we use ObfusLM on Llama3-8B to study this effect on the Alpaca~\cite{alpaca} dataset. Note that denoising is not used for ObfusLM, since it does not inject DP noise into the hidden embeddings.

Table~\ref{tab:ablation-denoise} highlights the importance of renormalization for attacking both DPNR (RoBERTa) and ObfusLM (Llama3-8B). With renormalization alone, the average ASR on ObfusLM increases from 0.517 to 0.562, and the ASR on DPNR also improves substantially across all $\epsilon$ values. Denoising becomes more beneficial at larger noise scales in DPNR; for example, it raises ASR from 0.316 to 0.349 when $\epsilon$ is 5. However, further reducing the PCA dimension does not always help: the gains from denoising are offset by the loss of information in overly aggressive dimensionality reduction. In our experiments, dimension of 256 yielded slightly worse performance. We leave the design and evaluation of more advanced denoising methods to future work.

\input{tables/ablation_ood}
\subsubsection{Out-of-domain attack}
In the experiments above, we followed the common setting in prior work~\cite{DBLP:conf/acl/LinYMZH0WLCD025, DBLP:conf/acl/ZhouLMGWDZZH23} and assumed that the shadow dataset is drawn from the same domain as the target data. We now relax this assumption and evaluate the attack in an out-of-domain setting. Here, the adversary may use any public corpus as shadow data for supervised training. Concretely, we construct the shadow dataset by randomly sampling 60,000 sentences from the English Wikipedia dump (20231101.en), and report representative results on ObfusLM ($\epsilon=0.1$), Santxt ($\epsilon=3$), TextObfuscator ($\epsilon=3$), and DPNR ($\epsilon=10$) in Table~\ref{tab:ablation-ood}.

Compared with the in-domain results (Table~\ref{tab:classification} and Figure~\ref{fig:dp}), both InvBert and \NAME\ are affected by distribution shift and exhibit a moderate drop in ASR. For embedding-DP schemes, \NAME\ is more robust than InvBert, as its unsupervised component can adapt to the target distribution. For ObfusLM, both attacks suffer a larger (relative) performance drop, likely due to two sources of shift: out-of-domain shadow data and the obfuscation gap induced by differing vocabulary permutations and clustering. 
Nonetheless, all evaluated schemes remain insecure, with recovery rates on most defenses well above 70\% and ObfusLM still close to 50\%. \revised{We note that for ObfusLM, the shadow models already use \emph{different} obfuscation configurations (different permutations and clustering seeds) from the target, so the OOD setting simultaneously tests both domain shift and configuration mismatch. A more controlled study that systematically varies the degree of configuration mismatch (e.g., number of clusters, DP variant) is left for future work.}

\revised{
\subsubsection{Effect of unlabeled data volume}\label{sec:ablation_volume}
A key assumption of \NAME\ is that the adversary can collect a sufficient volume of unlabeled obfuscated embeddings from the target deployment. To assess how attack performance depends on this quantity—and whether \NAME\ transfers to unseen held-out prompts—we vary the number of unlabeled samples while fixing the evaluation target to the SST-2 validation split (872 samples).

\begin{figure}[!t]
  \centering
  \includegraphics[width=0.95\linewidth]{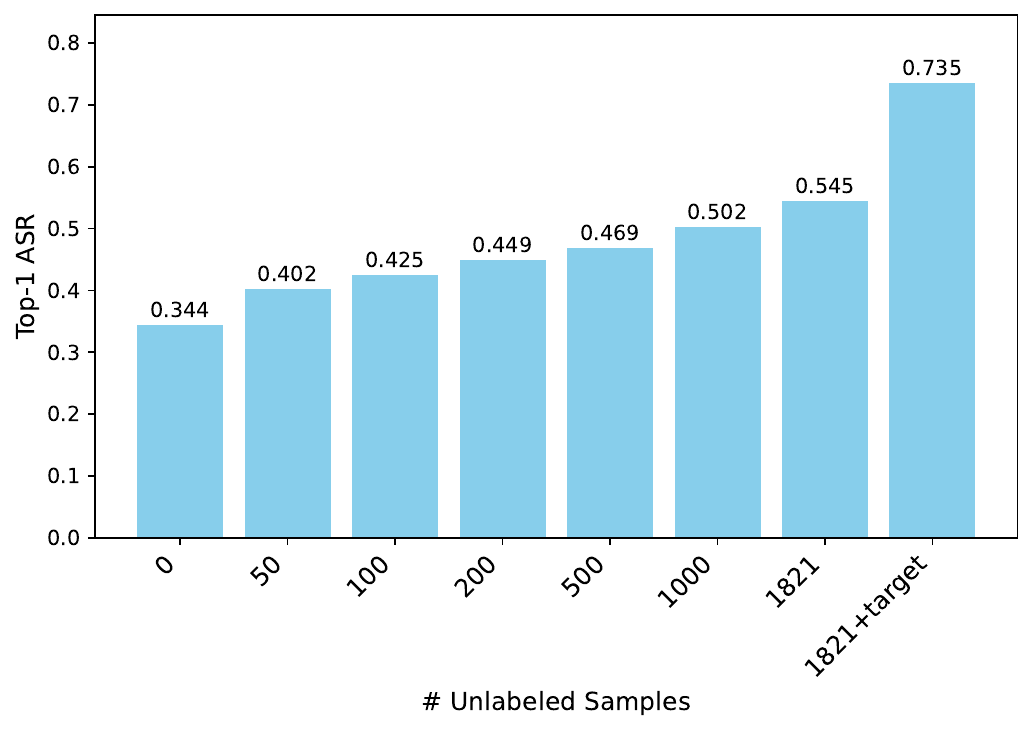}
  \caption{Effect of unlabeled data volume on top-1 ASR (ObfusLM, $\epsilon{=}0.1$, RoBERTa, SST-2). The x-axis labels denote the number of unlabeled samples used during training; ``1821+target'' includes the evaluation target in the unlabeled set (transductive setting). For all other bars, the evaluation target is excluded from the unlabeled pool, testing transfer to held-out prompts.}
  \label{fig:ablation_unlabeled}
\end{figure}

Figure~\ref{fig:ablation_unlabeled} shows the results on ObfusLM ($\epsilon{=}0.1$) with RoBERTa. With zero unlabeled samples, \NAME\ reverts to a purely supervised attack and achieves a top-1 ASR of 0.344, comparable to InvBert. As the unlabeled pool grows, ASR increases monotonically: even 50 unlabeled samples yield a noticeable improvement to 0.402, and the full test split (1{,}821 samples) raises ASR to 0.545. Crucially, in all of these configurations the evaluation target (the validation split) is \emph{not} included in the unlabeled training set, demonstrating that \NAME\ transfers to held-out prompts without requiring transductive access.

The rightmost bar (``1821+target'') corresponds to the transductive setting used in our main experiments, where the unlabeled set includes the evaluation target. ASR jumps to 0.735, a substantial gain over the non-transductive configuration with the same number of unlabeled samples (0.545). This gap reflects the benefit of the adversary's model adapting to embeddings drawn from the exact evaluation distribution. Nevertheless, the non-transductive results confirm that \NAME\ remains effective even when the adversary has no access to the evaluation prompts—addressing the concern that the attack relies solely on transductive leakage.
}

\subsection{Discussion} \label{sec:discussion}

\revised{
Our evaluation suggests that current obfuscation-based defenses face a \emph{task-dependent} tension in their privacy-utility trade-off against \NAME. On token-level tasks (e.g., NER) and generation tasks, the tension is acute: strong obfuscation can substantially degrade token-level information to the point of making defenses impractical, while weaker obfuscation preserves enough signal for an adaptive adversary to recover a substantial fraction of original tokens. On simpler sentence-level classification (e.g., SST-2), some DP-based defenses (CAPE, DP-Forward at low $\epsilon$) can simultaneously maintain task accuracy and limit token recovery, suggesting that the tension is most pronounced for tasks that require fine-grained token-level information. We view this not as a deficiency of any specific scheme, but as a structural challenge that future obfuscation-based designs will need to address, particularly for token-level and generation workloads.
}

Among existing alternatives, cryptographic approaches such as homomorphic encryption and secure multiparty computation~\cite{DBLP:conf/ndss/ZhangYH0LWH00025, DBLP:conf/ccs/MoonYJK25, DBLP:conf/ndss/LuHGL000WC25, DBLP:conf/sp/PangZMZS24} provide stronger guarantees that remain robust to attacks like \NAME\ while still supporting complex tasks, although they currently incur substantial computational and communication overhead. We refer readers to Appendix~\ref{sec:crypto} for representative performance results. We see our findings as motivation for continued research on lightweight obfuscation-based defenses, and on hybrid designs that combine obfuscation with cryptographic primitives to achieve a more practical balance between security and efficiency.

%% file: tables/classification.tex
\begin{table*}[t]
\centering

\resizebox{\linewidth}{!}{
\begin{threeparttable}
\caption{Attack performance on various defenses in classification task SST-2. \revised{Cell shading indicates ASR magnitude (darker = higher recovery); bold marks the best attack for each metric.}}\label{tab:classification}

\setlength{\tabcolsep}{1.5mm}
\begin{tabular}{l|l|c|ccc|ccc|ccc|ccc|ccc|ccc}
\toprule
\multirow{2}{*}{Model} 
&
\multirow{2}{*}{Defense} 
& 
\multirow{2}{*}{Acc.} 
& \multicolumn{3}{c|}{\textbf{KNN (ASR $\uparrow$)}} 
& \multicolumn{3}{c|}{\textbf{ER (ASR $\uparrow$)}} 
& \multicolumn{3}{c|}{\textbf{TBS (ASR $\uparrow$)}} 
& \multicolumn{3}{c|}{\textbf{DEML (ASR $\uparrow$)}} 
& \multicolumn{3}{c|}{\textbf{InvBert (ASR $\uparrow$)}} 
& \multicolumn{3}{c@{}}{\textbf{Ours (ASR $\uparrow$)}}
\\
& &
& Top-1 & RougeL & BLEU
& Top-1 & RougeL & BLEU
& Top-1 & RougeL & BLEU
& Top-1 & RougeL & BLEU
& Top-1 & RougeL & BLEU
& Top-1 & RougeL & BLEU
\\ \midrule

\multirow{8}{*}{\makecell{BERT\\$|V|\approx 30$K}} &

Plain$^\text{\textdagger}$
                        & 0.927
                        & \score{0.945} & \score{0.995} & \score{0.916}
                        & \best{1.000} & \best{1.000} & \best{1.000}
                        & \best{1.000} & \best{1.000} & \best{1.000}
                        & \best{1.000} & \best{1.000} & \best{1.000}
                        & \best{1.000} & \best{1.000} & \best{1.000}
                        & \best{1.000} & \best{1.000} & \best{1.000}
                        \\
&
SentinelLMs$^\text{\textdaggerdbl}$             & 0.935
                        & \score{0.950} & \score{0.995}  & \score{0.93}
                        & \best{1.000} & \best{1.000} & \best{1.000}
                        & \best{1.000} & \best{1.000} & \best{1.000}
                        & \best{1.000} & \best{1.000} & \best{1.000}
                        & \best{1.000} & \best{1.000} & \best{1.000}
                        & \best{1.000} & \best{1.000} & \best{1.000} 
                        \\
&
Datamix                 & 0.842
                        & \score{0.397} & \score{0.451} & \score{0.195}
                        & \score{0.593} & \score{0.560} & \score{0.373}
                        & \score{0.596} & \score{0.561} & \score{0.379}
                        & \score{0.587} & \score{0.559} & \score{0.356}
                        & \score{0.821} & \score{0.780} & \score{0.657}
                        & \best{0.927} & \best{0.916} & \best{0.844}
                        \\
&
Santext ($\epsilon:3$)  & 0.831
                        & \score{0.598} & \score{0.569} & \score{0.263}
                        & \score{0.615} & \score{0.563} & \score{0.275}
                        & \score{0.591} & \score{0.537} & \score{0.243}
                        & \score{0.612} & \score{0.554} & \score{0.262}
                        & \score{0.716} & \score{0.687} & \score{0.472}
                        & \best{0.775} & \best{0.751} & \best{0.553}
                        \\
&
Custext ($\epsilon: 3$)  & 0.892
                        & \score{0.845} & \score{0.854} & \score{0.681}
                        & \score{0.883} & \score{0.856} & \score{0.715}
                        & \score{0.884} & \score{0.857} & \score{0.717}
                        & \score{0.884} & \score{0.858} & \score{0.715}
                        & \score{0.911} & \score{0.894} & \score{0.787}
                        & \best{0.913} & \best{0.897} & \best{0.786}
                        \\
&
TextObf. ($\epsilon:3$) & 0.904
                        & \score{0.120} & \score{0.206} & \score{0}
                        & \score{0.613} & \score{0.627} & \score{0.253}
                        & \score{0.769} & \score{0.787} & \score{0.479}
                        & \score{0.631} & \score{0.661} & \score{0.238}
                        & \best{0.992} & \best{0.991} & \best{0.972}
                        & \score{0.991} & \best{0.991} & \score{0.970}
                        \\
&
Obf.LM ($\epsilon:3$)  & 0.892
                        & \score{0.357} & \score{0.438} & \score{0.062}
                        & \score{0.341} & \score{0.374} & \score{0.052}
                        & \score{0.375} & \score{0.389} & \score{0.072}
                        & \score{0.430} & \score{0.446} & \score{0.096}
                        & \score{0.671} & \score{0.644} & \score{0.355}
                        & \best{0.807} & \best{0.779} & \best{0.565}
                        \\
&                        
Obf.LM ($\epsilon:0.1$) & 0.851
                        & \score{0.207} & \score{0.246} & \score{0.016}
                        & \score{0.204} & \score{0.238} & \score{0.015}
                        & \score{0.204} & \score{0.239} & \score{0.015}
                        & \score{0.204} & \score{0.239} & \score{0.016}
                        & \score{0.281} & \score{0.260} & \score{0.041}
                        & \best{0.619} & \best{0.589} & \best{0.317}
                        \\

\midrule

\multirow{8}{*}{\makecell{RoBERTa\\ $|V|\approx 50$K}} &

Plain$^\text{\textdagger}$
                        & 0.954
                        & \score{0.901} & \score{0.975} & \score{0.767}
                        & \best{1.000} & \best{1.000} & \best{1.000}
                        & \best{1.000} & \best{1.000} & \best{1.000}
                        & \best{1.000} & \best{1.000} & \best{1.000}
                        & \best{1.000} & \score{0.999} & \score{0.999}
                        & \best{1.000} & \best{1.000} & \best{1.000}
                        \\
&
SentinelLMs$^\text{\textdaggerdbl}$             & 0.947
                        & \score{0.872} & \score{0.963} & \score{0.724}
                        & \best{1.000} & \best{1.000} & \best{1.000}
                        & \best{1.000} & \best{1.000} & \best{1.000}
                        & \score{0.989} & \score{0.997} & \score{0.967}
                        & \score{0.998} & \score{0.999} & \score{0.997}
                        & \best{1.000} & \best{1.000} & \best{1.000} 
                        \\
&
Datamix                 & 0.900
                        & \score{0.014} & \score{0.029} & \score{0}
                        & \score{0.648} & \score{0.597} & \score{0.459}
                        & \score{0.639} & \score{0.601} & \score{0.447}
                        & \score{0.529} & \score{0.535} & \score{0.291}
                        & \score{0.819} & \score{0.759} & \score{0.639}
                        & \best{0.938} & \best{0.926} & \best{0.882} 
                        \\
&
Santext ($\epsilon:3$)  & 0.846
                        & \score{0.566} & \score{0.621} & \score{0.253}
                        & \score{0.645} & \score{0.550} & \score{0.259}
                        & \score{0.633} & \score{0.541} & \score{0.255}
                        & \score{0.627} & \score{0.536} & \score{0.242}
                        & \score{0.762} & \score{0.721} & \score{0.503}
                        & \best{0.844} & \best{0.813} & \best{0.687}
                        \\
&
Custext ($\epsilon:3$)  & 0.930
                        & \score{0.811} & \score{0.874} & \score{0.571}
                        & \score{0.915} & \score{0.895} & \score{0.764}
                        & \score{0.915} & \score{0.894} & \score{0.766}
                        & \score{0.904} & \score{0.892} & \score{0.737}
                        & \best{0.943} & \score{0.929} & \best{0.846}
                        & \best{0.943} & \best{0.931} & \score{0.842}
                        \\
&
TextObf. ($\epsilon:3$) & 0.907
                        & \score{0} & \score{0} & \score{0}
                        & \score{0.002} & \score{0.015} & \score{0.002}
                        & \score{0.008} & \score{0.020} & \score{0.001}
                        & \score{0.005} & \score{0.015} & \score{0.001}
                        & \score{0.793} & \score{0.760} & \score{0.508}
                        & \best{0.916} & \best{0.907} & \best{0.762}  
                        \\
&
Obf.LM ($\epsilon:3$)  & 0.898
                        & \score{0.195} & \score{0.374} & \score{0.031}
                        & \score{0.234} & \score{0.298} & \score{0.029}
                        & \score{0.239} & \score{0.305} & \score{0.030}
                        & \score{0.418} & \score{0.462} & \score{0.111}
                        & \score{0.724} & \score{0.694} & \score{0.406}
                        & \best{0.827} & \best{0.794} & \best{0.578} \\
&                        
Obf.LM ($\epsilon:0.1$) & 0.864
                        & \score{0.175} & \score{0.209} & \score{0.013}
                        & \score{0.164} & \score{0.181} & \score{0.011}
                        & \score{0.164} & \score{0.181} & \score{0.011}
                        & \score{0.177} & \score{0.200} & \score{0.013}
                        & \score{0.262} & \score{0.266} & \score{0.040}
                        & \best{0.735} & \best{0.688} & \best{0.416}\\


\bottomrule
\end{tabular}
\begin{tablenotes}[flushleft]
\footnotesize
\item[] \textdagger: Plain refers to the original model that is not obfuscated.
\item[] \textdaggerdbl: SentinelLMs is fixed to reflect its open-source code according to this work~\cite{DBLP:conf/emnlp/LinZCHYLD24}.

\end{tablenotes}

\end{threeparttable}
}
\end{table*}

%% file: tables/generation.tex
\begin{table*}[t]
\small
\centering

\resizebox{0.95\linewidth}{!}{
\begin{threeparttable}
\caption{Attack performance on generation models in the medical QA and email datasets. \revised{S-Top1 denotes Sensitive Top-1, the top-1 recovery rate computed over sensitive tokens only. Cell shading indicates ASR magnitude (darker = higher recovery); bold marks the best attack for each metric.}\label{tab:generation}}

\setlength{\tabcolsep}{1.5mm}
\begin{tabular}{l|l|l|l|cccc|cccc|cccc}
\toprule
\multirow{2}{*}{\textbf{Task}} 
& \multirow{2}{*}{\textbf{Model}} 
& \multirow{2}{*}{\textbf{Defense}} 
& \multirow{2}{*}{Rouge1}
& \multicolumn{4}{c|}{\textbf{TBS (ASR $\uparrow$)}} 
& \multicolumn{4}{c|}{\textbf{InvBert (ASR $\uparrow$)}} 
& \multicolumn{4}{c@{}}{\textbf{Ours (ASR $\uparrow$)}} \\
& & &
& Top-1 & RougeL & BLEU & S-Top1
& Top-1 & RougeL & BLEU & S-Top1
& Top-1 & RougeL & BLEU & S-Top1
\\ \midrule

\multirow{10}{*}{\makecell{Medical QA}} 
& \multirow{5}{*}{\makecell{Llama3-8B\\$|V|\approx 128$K}} 
& Plain
& 0.804
& \score{0.974} & \score{0.986} & \score{0.910} & \score{0.970}
& \score{0.998} & \best{0.999} & \best{0.997} & \score{0.998}
& \best{0.999} & \best{0.999} & \best{0.997} & \best{0.999}
\\

& & Santext ($\epsilon: 3$) 
& 0.539
& \score{0.571} & \score{0.514} & \score{0.261} & \score{0.556}
& \score{0.663} & \score{0.669} & \score{0.371} & \score{0.638}
& \best{0.711} & \best{0.685} & \best{0.423} & \best{0.690}
\\

& & DPNR ($\epsilon: 15$) 
& 0.725
& \score{0.832}        & \score{0.821} & \score{0.544} & \score{0.845}
& \score{0.952}        & \best{0.935} & \best{0.856} & \score{0.952}
& \best{0.953} & \best{0.935} & \best{0.856} & \best{0.954}
\\

& & DP-Forward ($\epsilon: 15$) 
& 0.023 
& \score{0}     & \score{0.012} & \score{0.001} & \score{0}
& \score{0.080} & \score{0.136} & \score{0.008} & \score{0.135}
& \best{0.084} & \best{0.147} & \best{0.014} & \best{0.151}
\\

& & ObfusLM ($\epsilon: 1$) 
& 0.471
& \score{0.216} & \score{0.357} & \score{0.021} & \score{0.205}
& \score{0.457} & \score{0.510} & \score{0.148} & \score{0.345}
& \best{0.804} & \best{0.759} & \best{0.537} & \best{0.769}
\\

\cmidrule{2-16}

& \multirow{5}{*}{\makecell{Qwen3.5-27B\\$|V|\approx 248$K}} 
& Plain
& 0.96
& \score{0.984}        & \score{0.991}        & \score{0.920} & \score{0.984}
& \score{0.998}        & \best{0.999} & \best{0.999} & \best{0.999}
& \best{0.999} & \best{0.999} & \score{0.998} & \best{0.999}
\\

& & Santext ($\epsilon: 3$)
& 0.514
& \score{0.617} & \score{0.584} & \score{0.331} & \score{0.602}
& \score{0.676} & \score{0.679} & \score{0.409} & \score{0.647}
& \best{0.725} & \best{0.720} & \best{0.499} & \best{0.693}
\\
  
& & DPNR ($\epsilon: 15$) 
& 0.043
& \score{0}            & \score{0.010}        & \score{0.001} & \score{0}
& \score{0.273}        & \score{0.284}        & \score{0.098} & \score{0.324}
& \best{0.331} & \best{0.337} & \best{0.137} & \best{0.374}
\\

& & DP-Forward ($\epsilon: 15$) 
& 0
& \score{0}     & \score{0}     & \score{0} & \score{0}
& \best{0.083} & \score{0.147} & \score{0.014} & \score{0.140}
& \best{0.083} & \best{0.148} & \best{0.015} & \best{0.142}
\\

& & ObfusLM ($\epsilon: 1$) 
& 0.353
& \score{0}            & \score{0.015}        & \score{0.001} & \score{0}
& \score{0.575}        & \score{0.651}        & \score{0.248} & \score{0.556}
& \best{0.852} & \best{0.81} & \best{0.61} & \best{0.839}
\\

\midrule

\multirow{10}{*}{\makecell{Email}} 
& \multirow{5}{*}{\makecell{Llama3-8B\\$|V|\approx 128$K}} 
& Plain
& 0.781
& \score{0.958}        & \score{0.981}        & \score{0.927} & \score{0.951}
& \best{0.999} & \best{0.999} & \best{0.998} & \best{0.997}
& \score{0.995}        & \score{0.994}        & \score{0.995} & \score{0.996}
\\

& & Santext ($\epsilon: 3$)
& 0.427
& \score{0.573} & \score{0.502} & \score{0.249} & \score{0.561}
& \score{0.655} & \score{0.663} & \score{0.356} & \score{0.639}
& \best{0.683} & \best{0.688} & \best{0.401} & \best{0.671}
\\
  
& & DPNR ($\epsilon: 15$) 
& 0.701
& \score{0.837}        & \score{0.831}        & \score{0.578} & \score{0.838}
& \score{0.931}        & \score{0.914}        & \score{0.811} & \score{0.931}
& \best{0.944} & \best{0.929} & \best{0.846} & \best{0.946}
\\

& & DP-Forward ($\epsilon: 15$) 
& 0.052
& \score{0}     & \score{0.016}        & \best{0.002} & \score{0}
& \best{0.037} & \score{0.011}        & \score{0} & \score{0.037}
& \score{0.036} & \best{0.072} & \score{0} & \best{0.038}
\\

& & ObfusLM ($\epsilon: 1$) 
& 0.396
& \score{0.189}        & \score{0.378}        & \score{0.027} & \score{0.166}
& \score{0.428}        & \score{0.500}        & \score{0.153} & \score{0.312}
& \best{0.723} & \best{0.705} & \best{0.424} & \best{0.702}
\\

\cmidrule{2-16}

& \multirow{5}{*}{\makecell{Qwen3.5-27B\\$|V|\approx 248$K}} 
& Plain
& 0.791
& \score{0.954} & \score{0.956} & \score{0.899} & \score{0.950}
& \best{0.998} & \best{0.999} & \best{0.997} & \best{1.000}
& \best{0.998} & \best{0.999} & \best{0.997} & \best{1.000}
\\

& & Santext ($\epsilon: 3$)
& 0.445
& \score{0.582} & \score{0.538} & \score{0.287} & \score{0.573}
& \score{0.657} & \score{0.670} & \score{0.366} & \score{0.624}
& \best{0.691} & \best{0.703} & \best{0.432} & \best{0.673}
\\
  
& & DPNR ($\epsilon: 15$) 
& 0.046
& \score{0}            & \score{0.012}        & \score{0.001} & \score{0}
& \score{0.155}        & \score{0.160}        & \score{0.040} & \score{0.160}
& \best{0.301} & \best{0.267} & \best{0.160} & \best{0.306}
\\

& & DP-Forward ($\epsilon: 15$) 
& 0.001
& \score{0}            & \score{0.014}        & \best{0.001} & \score{0}
& \score{0.037}        & \score{0.063}        & \score{0} & \best{0.038}
& \best{0.038} & \best{0.071} & \score{0} & \best{0.038}
\\

& & ObfusLM ($\epsilon: 1$) 
& 0.441
& \score{0}            & \score{0.011}        & \score{0.001} & \score{0}
& \score{0.419}        & \score{0.595}        & \score{0.183} & \score{0.338}
& \best{0.788} & \best{0.808} & \best{0.584} & \best{0.701}
\\

\bottomrule
\end{tabular}
\begin{tablenotes}[flushleft]
\footnotesize
\item[] 

\end{tablenotes}

\end{threeparttable}
}
\end{table*}

%% file: tables/ablation_basic.tex







\begin{table}[!t]
\centering

\caption{Ablation study of soft labeling and masked views.}
\label{tab:ablation-basic}

\setlength{\tabcolsep}{2mm}
\begin{tabular}{@{}cc|c|c}
\toprule
\multirow{2}{*}{Masked} & \multirow{2}{*}{Soft} 
& 
\multicolumn{1}{c|}{\makecell{\textbf{ObfusLM}\\ ($\epsilon=0.1$, RoBERTa)}}
&
\multicolumn{1}{c@{}}{\makecell{\textbf{DPNR}\\ ($\epsilon=10$, RoBERTa)}} 
\\
&  &
Top-1 ASR &
Top-1 ASR
\\ 
\midrule

  &  & 
0.377 &
0.753
\\

& \checkmark &
0.393 &
0.760
\\

\checkmark  & & 
0.702 &
0.762
\\

\checkmark & \checkmark  & 
0.735 &
0.773
\\

\bottomrule
\end{tabular}



\end{table}

%% file: tables/ablation_denoise.tex







\begin{table}[!t]
\centering

\caption{Ablation study of renormalization and denoising.}
\label{tab:ablation-denoise}

\setlength{\tabcolsep}{1.5mm}
\begin{tabular}{@{}cc|c|c|c|c}
\toprule
\multirow{2}{*}{Renorm.} & \multirow{2}{*}{Denoise} 
& 
\multicolumn{1}{c|}{\makecell{\textbf{ObfusLM}\\ Alpaca, Llama3-8B}}
&
\multicolumn{3}{c@{}}{\makecell{\textbf{DPNR}\\ SST-2, RoBERTa}} 
\\
&  &
$\epsilon=0.1$ &
$\epsilon=5$   &
$\epsilon=10$  &
$\epsilon=15$
\\ 
\midrule

  &  & 
0.517 &
0.259 &
0.605 &
0.605 
\\

& \checkmark &
-     &
0.278 &
0.608 &
0.613
\\

\checkmark  & & 
0.562 &
0.316 &
0.741 &
0.866
\\

\checkmark & \checkmark  & 
-     &
0.349 &
0.785 &
0.869
\\

\bottomrule
\end{tabular}



\end{table}

%% file: tables/ablation_ood.tex
\begin{table}[!t]
\centering

\caption{Performance of out-of-domain attack with RoBERTa on the SST-2 task. The ASR drop from the in-domain attack is annotated in the subscript.}
\label{tab:ablation-ood}

\setlength{\tabcolsep}{2mm}
\begin{tabular}{@{}c|c|c}
\toprule
\multirow{1}{*}{Defense}
& 
\multicolumn{1}{c|}{\makecell{\textbf{InvBert}}}
&
\multicolumn{1}{c@{}}{\makecell{\textbf{Ours}}} 
\\ 
\midrule

ObfusLM ($\epsilon=0.1$)  & 
$0.184_{\downarrow 0.078}$ &
$0.471_{\downarrow 0.264}$
\\

Santext ($\epsilon=3$) & 
$0.672_{\downarrow 0.09}$ &
$0.733_{\downarrow 0.111}$
\\

TextObfus. ($\epsilon=3$) &
$0.63_{\downarrow 0.163}$ &
$0.838_{\downarrow 0.078}$
\\

DPNR ($\epsilon=10$) & 
$0.449_{\downarrow 0.186}$ &
$0.716_{\downarrow 0.057}$
\\

\bottomrule
\end{tabular}



\end{table}

%% file: sections/related.tex
\section{Related Work}

\noindent \textbf{Defenses.} Obfuscation-based defenses perturb representations while preserving semantic information for utility. Early works use heuristic techniques: Datamix~\cite{DBLP:conf/eccv/LiuWGZH20} applies mixup, SentinelLMs~\cite{DBLP:conf/aaai/MishraLD24} uses geometric transformations, and TextFusion~\cite{DBLP:conf/emnlp/ZhouLGMFWDCZH22} fuses word segments (unsuitable for token-level tasks). Token-level DP methods (Santext, Custext~\cite{DBLP:conf/acl/YueDWLSC21, DBLP:conf/acl/ChenMWCN0C23}) replace tokens with substitutes. Embedding-level DP defenses (TextObfuscator, DPNR, CAPE, DP-Forward~\cite{DBLP:conf/acl/ZhouLMGWDZZH23, DBLP:conf/emnlp/LyuHL20, DBLP:conf/emnlp/PlantGG21, DBLP:conf/ccs/DuYC0H023}) add noise to hidden representations with different normalization strategies. ObfusLM~\cite{DBLP:conf/acl/LinYMZH0WLCD025} combines vocabulary permutation, clustering, and $(k, \epsilon)$-DP. Recent works Aloepri~\cite{DBLP:journals/corr/abs-2603-01499} and Eguard~\cite{DBLP:conf/aaai/LiuYLWQR26} propose novel mechanisms but lack open-source implementations, making fair comparison difficult.

\noindent \textbf{Attacks.} A growing body of work demonstrates that embedding inversion can reconstruct text from obfuscated representations. Distance-based attacks such as KNN~\cite{DBLP:conf/cikm/QuKY0BN21} and EDNN~\cite{DBLP:conf/emnlp/LinZCHYLD24} infer tokens by comparing obfuscated embeddings to pretrained word embeddings, though EDNN is specifically designed to exploit SentinelLMs. Prompt embedding optimization methods (ER, TBS~\cite{DBLP:conf/uss/Dong00C0Z25}, DEML~\cite{DBLP:conf/sp/0004ZWXYLZ25}) formulate token recovery as gradient-based optimization, but lack structural robustness to obfuscation noise. Model-based inversion approaches train explicit inversion models: InvBERT~\cite{DBLP:journals/corr/abs-2109-10104} predicts token IDs from shadow data, while MLC~\cite{DBLP:conf/ccs/SongR20} performs multi-label classification. However, these supervised methods suffer from distribution shift between shadow and target data. ALGEN~\cite{DBLP:journals/corr/abs-2502-11308} demonstrates few-shot inversion but requires (plain, obfuscated) pairs unavailable in our setting.

Other recent works explore different attack surfaces: PLeak~\cite{DBLP:conf/ccs/0002Y0BC24} and Pape et al.~\cite{DBLP:conf/uss/PapeMES25} investigate system prompt leaking attacks and defenses. These works operate under fundamentally different threat models and technical assumptions compared to our embedding inversion setting, as they focus on extracting system instructions or bypassing safety filters rather than recovering user prompts from obfuscated embeddings.

%% file: sections/conclusion.tex
\section{Conclusion}
We introduced \NAME, a semi-supervised embedding inversion attack against obfuscation-based defenses for language model services. Across nine defenses, \revised{five} tasks, and four model architectures, \NAME\ outperforms prior attacks under each scheme's recommended privacy parameters: e.g., 73.5\% top-1 recovery on ObfusLM ($\epsilon{=}0.1$) vs.\ 26.2\%, and 85.2\% on ObfusLM ($\epsilon{=}1$) for Qwen3.5-27B medical QA. Defenses that resist \NAME\ do so only by destroying task utility. \revised{On token-level and generation tasks, this reveals a fundamental tension; on simpler tasks, some DP defenses can maintain both.} These findings motivate stronger, potentially cryptographic or hybrid, protection mechanisms.

%% file: sections/appendix.tex
\appendix

\subsection{Open Science}
All benchmark datasets and models used in this paper are open-source and made publicly accessible by their original contributors.

\subsection{PCA-Based Linear Denoising}\label{sec:app_pca}
\begin{lemma}
    Suppose $g \in \RR^H$ is a noise vector where each dimension is independent with mean 0 and variance $\sigma^2$, i.e., $g \sim (0, \sigma^2I_H)$. Let $z=PP^Tg$. Then $\frac{\EE\|z\|^2}{\EE\|g\|^2} = \frac{K}{H}$.
\end{lemma}
\begin{proof}
    We have $\EE\|g\|^2 = H\sigma^2$ and $P^TP=I_K$, and:
    \begin{align*}
        \EE\|z\|^2 & = \EE(g^TPP^TPP^Tg) = \EE(g^TPP^Tg) \\
        & = \EE[\mathrm{tr}(PP^Tgg^T)] = \mathrm{tr}(PP^T\EE[xx^T]) \\
        & = \mathrm{tr}(PP^T\sigma^2I_H) = \sigma^2\mathrm{tr}(PP^T) \\
        & = K\sigma^2
    \end{align*}
    The last step is because $PP^T$ is an orthogonal projection matrix and thus $\mathrm{tr}(PP^T) = \mathrm{rank}(PP^T) = K$. 
\end{proof}

\textbf{Signal-to-noise ratio improvement.} Let $h \in \RR^H$ be a clean embedding and $x = h + g$ the noisy version after DP perturbation. Define the signal energy $\EE\|h\|^2$ and noise energy $\EE\|g\|^2 = H\sigma^2$. After PCA projection onto the top $K$ components, the projected signal energy is $\EE\|P P^T h\|^2 \approx \sum_{i=1}^K \lambda_i$ (where $\lambda_i$ are the top $K$ eigenvalues of the covariance matrix of clean embeddings), while the projected noise energy becomes $K\sigma^2$. Let $\rho_K = \frac{\sum_{i=1}^K \lambda_i}{\sum_{i=1}^H \lambda_i}$ denote the fraction of total signal energy retained by the top $K$ components. Before PCA, the signal-to-noise ratio is $\mathrm{SNR}_{\text{pre}} = \frac{\EE\|h\|^2}{H\sigma^2}$; after PCA, $\mathrm{SNR}_{\text{post}} = \frac{\rho_K \cdot \EE\|h\|^2}{K\sigma^2} = \frac{H}{K}\,\rho_K \cdot \mathrm{SNR}_{\text{pre}}$. Since $\rho_K$ is typically close to 1 for large vocabularies (the embedding space is low-rank), the dominant factor is $H/K$. For our experiments with $H=768$ and $K=384$, the theoretical SNR gain is roughly $2\times$.

\subsection{Implementation Details}\label{sec:app_implementation}
We provide detailed hyperparameters for \NAME\ below. Unless otherwise stated, we set $(p_m, \sigma) = (0.2, 0.2)$ for weak augmentation views, and $(p_m, \sigma)=(0.4, 0.4)$ for strong augmentation views. The smoothing coefficient for the EMA teacher is set to 0.998. We set the start and end unsupervised ratios $(\tau_0, \tau_{n-1})=(0.3, 0.7)$ to guarantee a balanced budget for each component while emphasizing more unsupervised training towards the end. The KL divergence size $k$ is fixed to 500.

We train all models with a learning rate ($\eta$) of $1 \times 10^{-4}$ using the AdamW optimizer. The learning rate schedule is linear with $5\%$ warmup steps. The batch size $B$ is 64. The number of epochs is set to 25 when using a single shadow model, and increased to 50 for the multi-shadow setting with 7 shadows (necessary for attacking ObfusLM~\cite{DBLP:conf/acl/LinYMZH0WLCD025} to achieve optimal performance). The number of optimization steps per epoch $E$ is set to $|D_{\mathrm{shadow}}|/B$.

\input{tables/dataset}
\subsection{Dataset Details}
\label{sec:app_dataset}
We provide dataset statistics in Table~\ref{tab:dataset}. The ``real split'' is the full source of unlabeled embeddings available to the attacker; the ``real target split'' is the evaluation subset. Below we describe each dataset.

\noindent \textbf{SST-2~\cite{DBLP:conf/emnlp/SocherPWCMNP13}.} Movie review sentences with binary sentiment labels.

\noindent \textbf{CoNLL-2003~\cite{DBLP:conf/conll/SangM03}.} English NER benchmark with PER, ORG, LOC, MISC labels. We report entity-level F1 via $\mathsf{seqeval}$~\cite{seqeval}.

\noindent \textbf{AG News~\cite{DBLP:conf/nips/ZhangZL15}.} News headlines in four classes (World, Sports, Business, Sci/Tech).

\noindent \textbf{Medical Meadow MedQA~\cite{jin2020disease, medalpaca-medqa}.} Medical licensing exam questions with multiple-choice answers.

\noindent \textbf{Enron Emails~\cite{corbt-enron}.} Corporate emails from the Enron corpus. We merge it with \texttt{corbt/enron\_emails\_sample\_questions} on HuggingFace, pairing email excerpts with contextual questions to create instruction-following pairs for generation.


\input{tables/classification-multi}

\subsection{Experimental Results on Multi-Class Classification Tasks}\label{sec:classification-multi}
We omit SentinelLMs and the $\epsilon=3$ configuration of ObfusLM and report only the more informative multi-class classification results in Table~\ref{tab:classification-multi}. As before, we focus on RoBERTa-based models; the corresponding BERT results exhibit the same qualitative trends and lead to identical conclusions.

On AG News, which appears to be an easier sentence-level classification benchmark than SST-2, all defenses maintain a task accuracy above 0.9 under the same privacy parameter settings. The attack results largely mirror the binary classification case, with two notable exceptions. First, TextObfuscator is highly effective against the KNN baseline: by minimizing a clustering-style objective that pulls token embeddings toward cluster centroids, it pushes them away from their original positions and disrupts nearest-neighbor recovery. However, it remains vulnerable to \NAME, which attains an average ASR of 0.845. Second, \NAME~reaches an average ASR of 0.818 on ObfusLM, dramatically exceeding InvBert's 0.267. \emph{\NAME~is the only attack that achieves an average ASR above 0.8 against all defenses, demonstrating stable generalization across heterogeneous obfuscation mechanisms.}

The CoNLL-2003 benchmark yields quite different behavior, as token-level classification is substantially more sensitive to noise and token-level perturbations. Datamix and Santext fail to preserve utility on this task and suffer from low task accuracy, yet \NAME~still achieves high average ASR values of 0.941 and 0.774, respectively. For Santext, the performance gain over InvBert is modest, for the same reasons discussed earlier (token-level randomization limiting recoverability). For ObfusLM, \NAME's recovery rate is lower than on sentence-level tasks but still exceeds that of InvBert by a significant margin.


\subsection{Distribution of Hidden States and Noise}\label{sec:app_hidden_noise}

\begin{figure}[!t]
  \centering
  \subfloat[TextObfuscator $\epsilon=3.0$]{
	\includegraphics[width=0.49\linewidth]{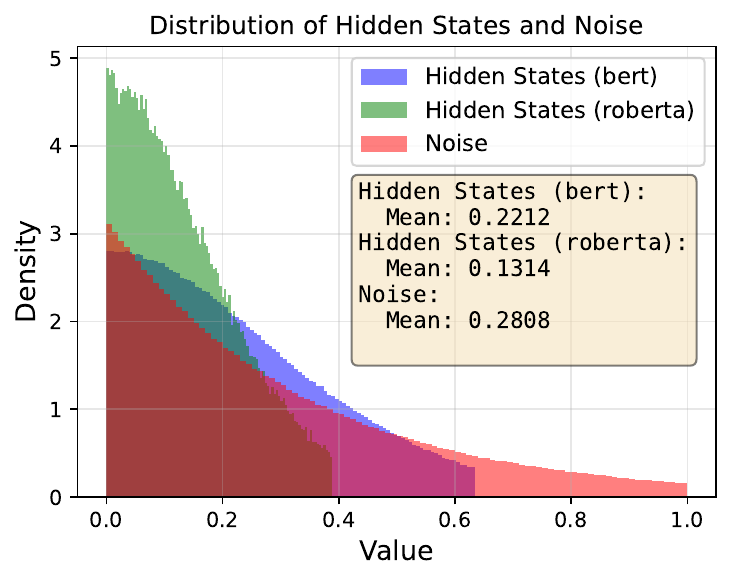}
  }
  \subfloat[DPNR $\epsilon=15.0$]{
	\includegraphics[width=0.49\linewidth]{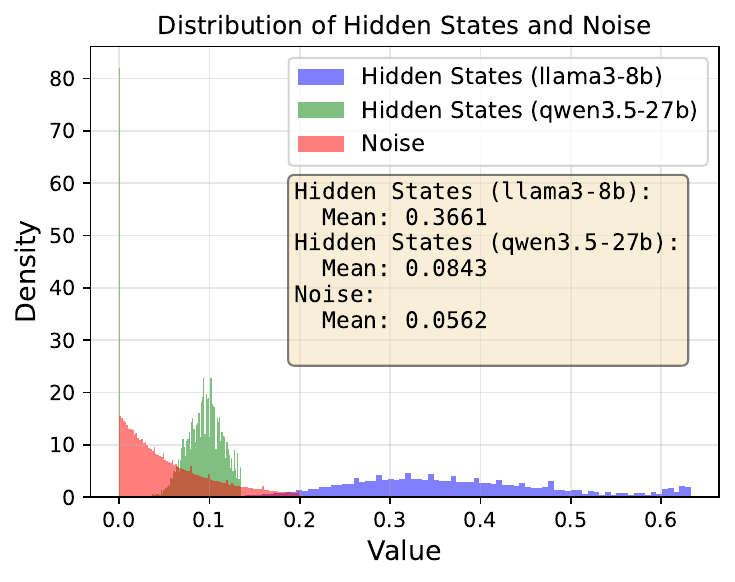}
  }
  \caption{Comparison of hidden state magnitudes and injected noise across different defenses and models. \textbf{(a)} TextObfuscator ($\epsilon=3.0$) on SST-2: The injected noise substantially exceeds the magnitude of RoBERTa's hidden states, explaining the degraded attack performance compared to BERT. \textbf{(b)} DPNR ($\epsilon=15.0$) on Medical QA: The noise dominates the hidden states of Qwen3.5-27B, resulting in lower attack success rates compared to the smaller LLaMA3-8B model. These observations highlight how model architecture and defense mechanisms interact to affect inversion vulnerability.}
  \label{fig:dist_hidden_noise}
\end{figure}

Figure~\ref{fig:dist_hidden_noise} illustrates why the same defense can lead to different inversion performance across architectures. Under DPNR ($\epsilon=15$), LLaMA3-8B hidden states have an absolute mean of 0.3661 while Qwen3.5-27B's are 0.0843; the injected noise (mean 0.0562) is model-independent. The resulting SNR is $\sim$6.5 for LLaMA3-8B but only $\sim$1.5 for Qwen3.5-27B, explaining why attack performance degrades on the latter. DPNR's min-max normalization lacks theoretical sensitivity guarantees, causing this model-dependent vulnerability.

Nonlinear inversion models exhibit inherent noise robustness through their layered architecture, while direct embedding optimization methods (ER, TBS, DEML) lack this resilience and are more sensitive to noise scale.

\input{tables/crypto}
\subsection{Performance of Cryptographic Solutions}\label{sec:crypto}
To illustrate the current status of cryptographic defenses, Table~\ref{tab:crypto} reports the cost of a single BERT inference (sequence length 128) for several state-of-the-art schemes. These approaches remain far too expensive for practical deployment.

\input{tables/atk_examples}
\subsection{Example Attacks on ObfusLM} \label{sec:app_atk_example}
Table~\ref{tab:atk_examples} presents illustrative attack examples against ObfusLM ($\epsilon=0.1$) configured with RoBERTa on SST-2. In most cases, InvBert fails to reconstruct fluent sentences that preserve the original semantics, whereas \NAME\ recovers the majority of the original content. 

\revised{
\subsection{Mixed Training: Motivation and Interpretation}\label{sec:app_mixed}
Equation~\ref{eq:joint} with a fixed $\lambda$ is standard joint optimization. However, when $\loss_s$ and $\loss_u^{(k)}$ have conflicting gradients, a fixed $\lambda$ yields a compromise update that attenuates useful components—a phenomenon known as \emph{gradient conflict}~\cite{DBLP:conf/nips/YuK0LHF20}. Empirically, no single $\lambda$ works well across all defenses and training stages.

\noindent\textbf{Alternating optimization.} Our mixed training alternates the two objectives: within each epoch $i$, we perform $(1-\tau_i)E$ steps on $\loss_s$ alone, then $\tau_i E$ steps on $\loss_u^{(k)}$ alone. This avoids per-step gradient interference while achieving an effect similar to joint optimization with epoch-wise weight $\lambda_i = \tau_i/(1-\tau_i)$, with two key differences:

\begin{enumerate}
    \item The two gradients never appear in the same update step, so per-step destructive interference is avoided. This shares a similar motivation with the projection step in PCGrad \cite{DBLP:conf/nips/YuK0LHF20}, but operates in the \emph{time-allocation} space rather than the gradient-direction space.
    \item The weighting is non-stationary: $\tau_i$ increases linearly from $\tau_0$ to $\tau_{n-1}$, allowing the model to first acquire enough supervised signal before the unsupervised objective begins to dominate. This resembles \emph{curriculum learning} over objectives \cite{DBLP:conf/icml/BengioLCW09, DBLP:journals/corr/abs-2101-10382}.
\end{enumerate}

We therefore view mixed training as a principled alternative to fixed-weight joint optimization. With a linear schedule, the average unsupervised ratio equals $(\tau_0+\tau_{n-1})/2$; we set $(\tau_0,\tau_{n-1})=(0.3,0.7)$, allocating roughly equal total budget to each component while shifting emphasis toward unsupervised learning in later epochs.
}






\revised{
\subsection{Sensitive Token Tagging Prompt}\label{sec:app_sensitive_tagging}

We use the following prompt to annotate sensitive tokens in the Medical QA and Email datasets via GPT-5.6:
}

\begin{framed}
\begin{Verbatim}[fontsize=\footnotesize, breaklines=true, breakanywhere=true, breakautoindent=false, breaksymbolleft={}]
You are a sensitive information detection agent. Your task is to identify
and annotate sensitive information in English text by wrapping each
sensitive span with <sensitive> and </sensitive> tags.

## Sensitive Information Categories

Mark the following types of information as sensitive:

1. Personally Identifiable Information (PII)
   - Full names, partial names with titles (e.g., "Dr. Smith", "Mr. John")
   - Phone numbers, email addresses, physical addresses
   - Government IDs: SSN, passport numbers, driver's license numbers
   - Date of birth, age combined with gender or other identifying details

2. Medical & Health Information (PHI)
   - Diagnoses and psychiatric conditions (e.g., "major depressive disorder")
   - Prescription drugs and dosages (e.g., "fluoxetine", "20mg Lisinopril")
   - Vital signs and lab values (e.g., "temperature is 38.9 C",
     "blood pressure is 152/94 mm Hg")
   - Physical examination findings (e.g., "Patellar reflexes are 4+
     bilaterally")
   - Any specific clinical observations tied to an individual

3. Financial Information
   - Credit card numbers, bank account numbers
   - Transaction amounts, salary figures
   - Insurance policy numbers

4. Credentials & Secrets
   - Passwords, API keys, access tokens
   - Private keys, secret questions and answers

5. Confidential & Proprietary Information
   - Internal project codenames, unpublished research data
   - Trade secrets, proprietary algorithms
   - Confidential business strategies or internal metrics

6. Location Information
   - Specific office locations, GPS coordinates
   - Private residential addresses

## Annotation Rules

- Wrap each sensitive span inline within the original text. Do not
  rephrase, reorder, or summarize.
- Tag the minimal meaningful span. For example, tag "blood pressure is
  152/94 mm Hg" as a unit, not individual numbers separately.
- Do NOT tag generic symptoms in isolation (e.g., "nausea", "headache")
  unless they appear alongside identifiable clinical context.
- Do NOT tag common words, articles, or structural text.
- If no sensitive information is found, return the text unchanged.
- Preserve all original formatting, line breaks, and punctuation.

## Output Format

Return the full text with sensitive spans wrapped in tags. Example:

Input:
  The patient John Smith was prescribed 20mg Lisinopril and can be
  reached at 555-123-4567.

Output:
  The patient <sensitive>John Smith</sensitive> was prescribed
  <sensitive>20mg Lisinopril</sensitive> and can be reached at
  <sensitive>555-123-4567</sensitive>.

## Input

{text}

## Output
\end{Verbatim}
\end{framed}


%% file: tables/dataset.tex
\begin{table}[!t]
\centering

\caption{Datasets}
\label{tab:dataset}

\begin{tabular}{@{}l|c|c|c}
\toprule
Dataset & Shadow & Real & Real Target
\\
\midrule
SST-2       & 67349  & 2693  & 872
\\
AG News     & 120000 & 7600  & 7600
\\
CoNLL-2003  & 14041  & 6703  & 3250
\\
MedQA       & 8142  & 2036  & 2036
\\
Email       & 4000 & 1000 & 1000
\\
Alpaca      & 41408  & 10352 & 10352
\\

\bottomrule
\end{tabular}

\end{table}

%% file: tables/classification-multi.tex
\begin{table*}[t]
\centering

\caption{Attack performance on various defenses of sentence-level and token-level multi-class classification tasks.}\label{tab:classification-multi}

\setlength{\tabcolsep}{1.5mm}
\begin{tabular}{l|l|c|cccc|cccc|cccc}
\toprule
\multirow{2}{*}{Task} 
& \multirow{2}{*}{Defense} 
& \multirow{2}{*}{Acc./F1} 
& \multicolumn{4}{c|}{\textbf{KNN (ASR $\uparrow$)}} 
& \multicolumn{4}{c|}{\textbf{InvBert (ASR $\uparrow$)}} 
& \multicolumn{4}{c@{}}{\textbf{Ours (ASR $\uparrow$)}} \\
& &
& Top-1 & Top-5 & RougeL & Avg.
& Top-1 & Top-5 & RougeL & Avg.
& Top-1 & Top-5 & RougeL & Avg.
\\ \midrule

\multirow{6}{*}{\makecell{AG News\\ (Sentence)}} &
Plain
                        & 0.949
                        & 0.881   & 0.998   & 0.957 & 0.945
                        & 1.000   & 1.000   & 1.000 & 1.000
                        & 1.000   & 1.000   & 1.000 & 1.000
                        \\
&
Datamix                 & 0.941
                        & 0.036   & 0.430   & 0.075 & 0.180
                        & 0.913   & 0.994   & 0.918 & 0.942
                        & 0.979   & 0.998   & 0.976 & 0.984  
                        \\
&
Santext ($\epsilon=3$)  & 0.915
                        & 0.544   & 0.614   & 0.600 & 0.586   
                        & 0.778   & 0.848   & 0.747 & 0.791  
                        & 0.804   & 0.876   & 0.773 & 0.818  \\
&
Custext ($\epsilon=3$)  & 0.941
                        & 0.765   & 0.870   & 0.819 & 0.818    
                        & 0.950   & 0.979   & 0.939 & 0.956
                        & 0.965   & 0.987   & 0.958 & 0.970 
                        \\
&
TextObf. ($\epsilon=3$) & 0.943
                        & 0.000   & 0.086   & 0.000 & 0.029
                        & 0.634   & 0.828   & 0.611 & 0.691 
                        & 0.811   & 0.933   & 0.790 & 0.845 
                        \\
&
ObfusLM ($\epsilon=0.1$)& 0.929
                        & 0.136   & 0.317   & 0.211 & 0.221
                        & 0.175   & 0.388   & 0.239 & 0.267 
                        & 0.781   & 0.924   & 0.750 & 0.818
                        \\

\midrule

\multirow{6}{*}{\makecell{CoNLL2003\\ (Token)}} &
Plain
                        & 0.955
                        & 0.751   & 0.984   & 0.838 & 0.858
                        & 1.000   & 1.000   & 0.997 & 0.999 
                        & 1.000   & 1.000   & 0.997 & 0.999
                        \\
&
Datamix                 & 0.004
                        & 0.011   & 0.475   & 0.023 & 0.17
                        & 0.784   & 0.980   & 0.705 & 0.823  
                        & 0.929   & 0.993   & 0.900 & 0.941
                        \\
&
Santext ($\epsilon=3$)  & 0.773
                        & 0.497   & 0.619   & 0.544 & 0.553   
                        & 0.743   & 0.811   & 0.682 & 0.745
                        & 0.768   & 0.837   & 0.716 & 0.774
                        \\
&
Custext ($\epsilon=3$)  & 0.929
                        & 0.676   & 0.874   & 0.737 & 0.762
                        & 0.920   & 0.951   & 0.892 & 0.921 
                        & 0.932   & 0.960   & 0.911 & 0.934
                        \\
&
TextObf. ($\epsilon=3$) & 0.931
                        & 0.033   & 0.334   & 0.067 & 0.145   
                        & 0.932   & 0.986   & 0.926 & 0.948 
                        & 0.957   & 0.992   & 0.951 & 0.967
                        \\

&
ObfusLM ($\epsilon=0.1$)& 0.862
                        & 0.133   & 0.298   & 0.155 & 0.195
                        & 0.277   & 0.534   & 0.262 & 0.358
                        & 0.563   & 0.803   & 0.484 & 0.644   \\

\bottomrule
\end{tabular}


\end{table*}

%% file: tables/crypto.tex
\begin{table}[!t]
\centering

\caption{Inference costs of cryptographic defenses.}
\label{tab:crypto}

\setlength{\tabcolsep}{2mm}
\begin{tabular}{@{}l|c|c}
\toprule
\multirow{1}{*}{Defense}
& 
Time (min, LAN)
&
Comm. (GB)
\\ 
\midrule
BOLT~\cite{DBLP:conf/sp/PangZMZS24} (CPU $\times$ 64) &
3 & 59 
\\
BumbleBee~\cite{DBLP:conf/ndss/LuHGL000WC25} (CPU $\times$ 64) &
3 & 6
\\
NEXUS~\cite{DBLP:conf/ndss/ZhangYH0LWH00025} (A100 $\times$ 4) &
0.6 & 0.16
\\
THOR~\cite{DBLP:conf/ccs/MoonYJK25} (A100 $\times$ 1) &
10 & -
\\

\bottomrule
\end{tabular}



\end{table}

%% file: tables/atk_examples.tex
\begin{table*}[!t]
\centering

\caption{Examples of Attacking Results.}
\label{tab:atk_examples}

\setlength{\tabcolsep}{2mm}
\begin{tabular}{@{}l|l|}
\toprule
\multirow{1}{*}{Method}
& 
Text
\\ 
\midrule
Original &
\red{if you 're hard up for raunchy college humor , this is your ticket right here .}
\\
InvBert &
have \red{you} 'ree approach compared to unbishtelymp questionses as i \red{your ticket}s tremERE :
\\
\NAME\ &
\red{if you 're hard up for raunchy college humor , this is your ticket right} there \red{.}
\\
\midrule
Original &
\red{there is n't nearly enough fun here , despite the presence of some appealing ingredients .}
\\
InvBert &
 '\red{there} i were\red{n't} exactly inferior SUNERE ; except. appearance for if desirable substance :  '
\\
\NAME\ &
\red{there is n't nearly enough fun here , despite the} efforts \red{of some appealing} material .
\\
\midrule
Original &
\red{while the ideas about techno-saturation are far from novel , they 're presented with a wry dark humor .}
\\
InvBert &
 hadplus. sensibilities \red{about} ex/s veins had cursby literature ; a 'thy offering as has \red{wry}s \red{dark}er probing :  had
\\
\NAME\ &
\red{while the ideas} of ex-s liberation \red{are far from} poetry \red{, they 're presented with a wry dark humor .}
\\

\bottomrule
\end{tabular}



\end{table*}